\documentclass[a4paper,11pt]{article}
\pdfoutput=1 

\usepackage{jcappub} 

\usepackage[T1]{fontenc}
\usepackage{mathrsfs}
\usepackage{amsmath}
\usepackage{upgreek}
\usepackage{multirow}
\usepackage{array}

\title{\boldmath \texttt{SpinOctopus}: A Versatile, User-Friendly, and Extensible Public Code for General-Relativistic Ray-Tracing in axisymmetric spacetime}

\author[a,b,1]{Shiyang Hu,\note{Corresponding author.}}
\author[a,b]{Dan Li,}
\author[c]{Chen Deng,}
\author[d]{Lina Zhang,}
\author[e]{and Wenfu Cao}

\affiliation[a]{School of Mathematics and Physics, University of South China, \\ Hengyang, 421001, People's Republic of China}
\affiliation[b]{Hunan Provincial Key Laboratory of Mathematical Modeling and Scientific Computing, University of South China, \\ Hengyang, 421001, People's Republic of China}
\affiliation[c]{School of Astronomy and Space Science, Nanjing University, \\ Nanjing, 210023, People's Republic of China}
\affiliation[d]{College of Science, Hunan Institute of Technology, \\ Hengyang, 421002, People's Republic of China}
\affiliation[e]{School of Physics and Technology, University of Jinan, \\ Jinan, 250022, People's Republic of China}

\emailAdd{husy$\_$arcturus@163.com}
\emailAdd{danli@usc.edu.cn}
\emailAdd{dengchen@nju.edu.cn}
\emailAdd{linazhang@hnit.edu.cn}
\emailAdd{wfcao0622@163.com}

\abstract{Relativistic ray-tracing algorithms serve as indispensable numerical tools for investigating gravitational theories and high-energy astrophysics. In this paper, we develop and release \texttt{SpinOctopus}, a versatile relativistic ray-tracing code tailored for arbitrary axisymmetric spacetimes, powered by a Fortran/Python hybrid architecture and OpenMP multi-threading acceleration. This numerical toolkit is capable of simulating timelike orbits and their corresponding light curves, null geodesics, as well as black hole images illuminated by a wide array of phenomenological sources, including geometrically thin and thick accretion disks, geometrically thin tilted disks, relativistic jets, spherically symmetric accretion flows, and point sources. To execute these tasks, users are solely required to specify the covariant metric tensor of the target spacetime and the simulation parameters within a dedicated configuration file. Benefiting from our customized adaptive step-size strategy for geodesic integration, \texttt{SpinOctopus} achieves an impressive processing efficiency of $\sim 800$ rays per second per thread on an Intel Core i7-14700 platform when simulating geometrically thick accretion disk images. Furthermore, the code provides transparent interfaces, allowing users to seamlessly modify the geometric and kinematic properties of the light sources. Additional capabilities, such as simulating Fe K${\alpha}$ fluorescent line profiles and geometrically thick tilted accretion disk images, will be introduced in future releases of the framework.}
\begin{document}
\maketitle
\flushbottom
\section{Introduction}
Using Very Long Baseline Interferometry (VLBI), the Event Horizon Telescope (EHT) collaboration has captured images of the supermassive black holes M87$^{*}$ and Sagittarius A$^{*}$ (Sgr A$^{*}$) \cite{2019ApJ...875L...1E,2022ApJ...930L..12E,2024A&A...681A..79E}. This scientific breakthrough not only answers the historical hand-drawn impression of a black hole from 1979 \cite{1979A&A....75..228L} and the outlook for black hole observations at the turn of the millennium \cite{2000ApJ...528L..13F}, but also opens up a new and effective window for the study of gravitational theories and high-energy astrophysics \cite{2020PhRvD.102l4004G,2023ApJ...942...47Y,2026A&A...705A.166S,2026A&A...705A.156S}. Consequently, it has triggered a surge of interest in numerically simulating black hole shadows \cite{2019PhRvD.100b4018G,2019ApJ...885L..33N,2020EPJC...80..872Z,Li:2021riw,Zeng:2021mok,Li:2021ypw,He:2022yse,Wu:2024juj,Peng:2020wun,Zeng:2022pvb,Hu:2022lek,Hu:2023bzy,Hu:2023pyd,Hu:2024imp,Hu:2026rqa,Chatterjee:2020eqc,White:2020cos,Li:2025jfq,Cunha:2016bpi,Cunha:2015yba,Tang:2022uwi,Wang:2021ara,Cao:2023ppv,Meng:2025ivb,Meng:2023htc,Zeng:2023fqy,2021ApJ...918....6C,2026ForPh..7470062L,Hou:2022eev,Zhang:2023bzv,Li:2024ctu,Zeng:2025tji,Wan:2026xzs,Huang:2025xqd,Huang:2024gtu,2021ApJ...920..155B,Wei:2026zwu,Chen:2025pgg,Wang:2022yvi,Guo:2022nto}, polarimetric images \cite{Gelles:2021kti,Qin:2022kaf,Zhang:2021hit,Shi:2024bpm,Wang:2025btn,Zhang:2023cuw,Zhou:2025moa,Yang:2025byw,Hou:2024qqo,Zeng:2026ntc,Aslam:2025hgl,Wang:2025qpv,Zeng:2025kyv,Yin:2025rao,Qin:2023nog,Li:2026lca}, and hotspot images \cite{Huang:2024wpj,Zhu:2025jqh,Zhu:2024vxw,Zhao:2025ouq,Wang:2025dyw,Hu:2022sej,Lee:2022rtg,Chen:2024jkm,Ruales:2026xjb,Hu:2025hdr}.

At the theoretical level, simulating a black hole image involves computing the electromagnetic radiation emitted by the accreting matter (the source) and projecting it onto the observer's screen along null geodesics. This process requires a class of post-processing numerical tools, whose underlying numerical strategy is dictated by the form of the source they process.

When the source is drawn from the grid data of a general-relativistic magnetohydrodynamic (GRMHD) simulation \cite{2019ApJS..243...26P,Li:2026xpk}, in which the radiation parameters and the dynamical quantities of the plasma near the black hole are specified as discrete three-dimensional data, the corresponding post-processing tool is usually termed a general-relativistic radiative transfer (GRRT) algorithm \cite{2020ApJ...897..148G,2018MNRAS.475...43M,2025ApJ...995...56N,Chan:2013tsa,Chan:2017igo,Bronzwaer:2018lde,Bronzwaer:2020kle,Cardenas-Avendano:2022csp,Huang:2024bar,Pelle:2022phf,2022ApJS..262...28W,Younsi:2012ha,Dexter:2016cdk,Pu:2016eml,Sharma:2023nbk,Pihajoki:2018ihj,2017MNRAS.464.4567T,Kawashima:2021icd,2016MNRAS.457..608N,2026arXiv260926298Z,Zhou:2026sew}. Conversely, when the emission and dynamics of the accreting matter are expressed analytically, the corresponding method is termed a general-relativistic ray-tracing algorithm. This approach can be further developed into a comprehensive numerical framework capable of simulating black hole images, computing energy spectra and light curves, and modeling polarization transport. Notably, both methods solve the radiative transfer equation and integrate null geodesics; they differ only in how the emission, absorption, and source dynamics are specified. Consequently, relativistic ray-tracing algorithms for analytical sources possess intrinsic advantages in code architecture, computational efficiency, and extensibility, having evolved into a substantial family of codes \cite{Cunha:2016bpi,2009ApJ...696.1616D,Vincent:2011wz,Bambi:2012tg,Psaltis:2010ww,Baubock:2011ke,2014CoPhC.185.2301M,2018GReGr..50...75Q,Riazuelo:2015shp,2022A&A...665A.139S,M:2022pex,Hu:2020usx,Zhong:2021mty,Davelaar:2021gxx,deLeon:2024exi,Zhou:2024dbc,2013ApJS..207....6Y,2014A&A...561A.127Y,Chen:2015cha,Garnier:2023lph,Bohn:2014xxa,2020PhRvD.101d4031G,Guo:2024mij}. Nevertheless, developing a ray-tracing algorithm that is simultaneously applicable to arbitrary axisymmetric spacetimes, sufficiently user-friendly, readily extensible with new sources and functionalities, and highly computationally efficient remains an outstanding challenge. The present work is motivated by these requirements.

In this paper, we develop \texttt{SpinOctopus}, a ray-tracing code applicable to arbitrary axisymmetric spacetimes. Built on Fortran and Python and integrated with the OpenMP (OMP) acceleration framework, the code is capable of simulating black hole critical curves, timelike particle orbits, and light curves, as well as rendering black hole images against the backgrounds of geometrically thin or thick accretion disks, jets, and point sources. Owing to an optimized light ray integration strategy, \texttt{SpinOctopus} computes light trajectories at a rate of approximately 800 rays per thread per second on an Intel Core i7-14700 processor. Regarding usability, users simply need to define the covariant metric tensor of the target spacetime and the required task parameters within a Python script; the code then automatically generates the corresponding Fortran model files to be invoked by the core modules. This streamlined input process ensures exceptional ease of use, eliminating the need to modify the Fortran numerical core when switching metrics or simulation tasks.

The remainder of this paper is organized as follows. In section 2, we describe the mathematical formalism underlying \texttt{SpinOctopus}, including the coordinate transformations between the local frame of the curved spacetime and the observer's local frame, the equations of motion for light rays, and the radiative transfer equation that yields the specific intensity of the rays. In section 3, taking the Kerr black hole as a representative example, we detail the integrated light sources, diverse functionalities, and computational efficiency of the code. In the final section, we present our conclusions and discussions. Throughout this paper, we adopt geometrized units, in which the gravitational constant, the speed of light, and the black hole mass are all set to unity ($G=c=M=1$). In certain sections, we also restore the units to the International System of Units (SI) for clarity.
\section{Ray-Tracing and Radiative Transfer}
\subsection{Metric and geodesics}
\texttt{SpinOctopus} is applicable to asymptotically flat, stationary, and axisymmetric spacetimes.\footnote{If the spacetime does not satisfy the asymptotic flatness condition, \texttt{SpinOctopus} can still execute the computations mathematically; however, the results may incur inaccuracies, as the coordinate transformations for the initial ray conditions rely on this assumption. Furthermore, our current framework is inapplicable to certain wormhole models, particularly traversable wormholes, as these scenarios necessitate specialized computational treatments for null geodesics.} In Boyer-Lindquist coordinates $x^{\mu}=\left(t,r,\theta,\varphi\right)$, the line element of the target spacetime is written as
\begin{equation}\label{1}
\textrm{d}s^{2} = g_{tt}\textrm{d}t^{2} + g_{rr}\textrm{d}r^{2} + g_{\theta\theta}\textrm{d}\theta^{2} + g_{\varphi\varphi}\textrm{d}\varphi^{2} + 2g_{t\varphi}\textrm{d}t\textrm{d}\varphi,
\end{equation}
where $g_{\mu\nu}$ denotes the covariant metric tensor of the spacetime. Letting $\Delta\left(r\right)$ denote the denominator of $g_{rr}$, the event horizon radius $r_{\textrm{eh}}$ of the central compact object can be determined by solving $\Delta\left(r_{\textrm{eh}}\right) = 0$. Under a preset iteration tolerance of $10^{-15}$, \texttt{SpinOctopus} employs the Newton's method to solve $\Delta\left(r_{\textrm{eh}}\right) = 0$, with the iteration formula expressed as
\begin{equation}\label{2}
r_{\textrm{eh}}^{n} = r_{\textrm{eh}}^{n-1} - \frac{\Delta\left(r_{\textrm{eh}}^{n-1}\right)}{\Delta^{\prime}\left(r_{\textrm{eh}}^{n-1}\right)}.
\end{equation}
Here, the prime denotes the derivative with respect to $r$, while $n$ and $n-1$ indicate consecutive iteration steps. Notably, the Newton's method requires an initial guess. To ensure the generality of the algorithm, \texttt{SpinOctopus} first scans $\Delta\left(r_{\textrm{eh}}\right) = 0$ to secure a rough root, and subsequently uses this root as the initial guess for the iteration to converge upon an accurate solution.

Based on the spacetime line element \eqref{1}, the Lagrangian $\mathcal{L}$ governing the particle motion is given by
\begin{equation}\label{3}
2\mathcal{L} = g_{\mu\nu}\dot{x}^{\mu}\dot{x}^{\nu} = g_{tt}\dot{t}^{2} + g_{rr}\dot{r}^{2} + g_{\theta\theta}\dot{\theta}^{2} + g_{\varphi\varphi}\dot{\varphi}^{2} + 2g_{t\varphi}\dot{t}\dot{\varphi},
\end{equation}
where $\dot{x}^{\mu} = \left(\dot{t},\dot{r},\dot{\theta},\dot{\varphi}\right)$ denotes the tangent vector (particle velocity) to the trajectory, defined as the derivative of the generalized coordinates $x^{\mu}$ with respect to the proper time $\tau$ (for massive particles) or the affine parameter $\lambda$ (for photons). Timelike particles and photons are subject to different Lagrangian constraints: the former satisfy $\mathcal{L}=-1/2$, whereas the latter require $\mathcal{L}=0$.

Following the Euler-Lagrange equations, we introduce the conjugate momenta of the particle, $p_{\mu} = \partial\mathcal{L}/\partial\dot{x}^{\mu}$, which are explicitly written as
\begin{eqnarray}
p_{t} &=& g_{tt}\dot{t} + g_{t\varphi}\dot{\varphi}, \label{4} \\
p_{r} &=& g_{rr}\dot{r}, \label{5} \\
p_{\theta} &=& g_{\theta\theta}\dot{\theta}, \label{6} \\
p_{\varphi} &=& g_{\varphi\varphi}\dot{\varphi} + g_{t\varphi}\dot{t}. \label{7}
\end{eqnarray}
Given that the spacetime possesses time-translational and rotational invariances---that is, the line element \eqref{1} does not explicitly depend on the coordinate time $t$ and the azimuthal angle $\varphi$---the particle motion admits two constants of motion: the specific energy $\mathcal{E}=-p_{t}$ and the specific angular momentum $\mathcal{P}=p_{\varphi}$.

Via the Legendre transformation, the Hamiltonian $\mathcal{H}$ corresponding to the Lagrangian $\mathcal{L}$ is obtained as
\begin{eqnarray}\label{8}
\mathcal{H} &=& p_{\mu}\dot{x}^{\mu} - \mathcal{L} = \frac{1}{2}g^{\mu\nu}p_{\mu}p_{\nu} \nonumber \\
 &=& \frac{1}{2}\left(g^{tt}p_{t}^{2} + g^{rr}p_{r}^{2} + g^{\theta\theta}p_{\theta}^{2} + g^{\varphi\varphi}p_{\varphi}^{2} + 2g^{t\varphi}p_{t}p_{\varphi}\right).
\end{eqnarray}
Here, $g^{\mu\nu}$ denotes the contravariant metric tensor. Analogous to the Lagrangian formulation, the Hamiltonian constraint requires $\mathcal{H}=-1/2$ for massive particles and $\mathcal{H}=0$ for photons. Evaluating the canonical equations $\dot{x}^{\mu}=\partial\mathcal{H}/\partial p_{\mu}$ and $\dot{p}_{\mu}=-\partial\mathcal{H}/\partial x^{\mu}$ yields the system of differential equations governing the particle motion:
\begin{eqnarray}
\dot{t} &=& g^{tt}p_{t} + g^{t\varphi}p_{\varphi}, \label{9} \\
\dot{r} &=& g^{rr}p_{r}, \label{10} \\
\dot{\theta} &=& g^{\theta\theta}p_{\theta}, \label{11} \\
\dot{\varphi} &=& g^{\varphi\varphi}p_{\varphi} + g^{t\varphi}p_{t}, \label{12} \\
\dot{p}_{t} &=& 0, \label{13} \\
\dot{p}_{r} &=& -\frac{1}{2}\left(g^{tt,r}p_{t}^{2} + g^{rr,r}p_{r}^{2} + g^{\theta\theta,r}p_{\theta}^{2} + g^{\varphi\varphi,r}p_{\varphi}^{2} + 2g^{t\varphi,r}p_{t}p_{\varphi}\right), \label{14} \\
\dot{p}_{\theta} &=& -\frac{1}{2}\left(g^{tt,\theta}p_{t}^{2} + g^{rr,\theta}p_{r}^{2} + g^{\theta\theta,\theta}p_{\theta}^{2} + g^{\varphi\varphi,\theta}p_{\varphi}^{2} + 2g^{t\varphi,\theta}p_{t}p_{\varphi}\right), \label{15} \\
\dot{p}_{\varphi} &=& 0. \label{16}
\end{eqnarray}
Here, $g^{\mu\nu,r}$ and $g^{\mu\nu,\theta}$ denote the partial derivatives of the contravariant metric components $g^{\mu\nu}$ with respect to $r$ and $\theta$, respectively.

It is evident that as long as the spacetime line element can be cast in the form of equation \eqref{1}, the equations of motion for both timelike particles and photons are universally governed by \eqref{9}---\eqref{16}. This universality underpins the broad compatibility of \texttt{SpinOctopus}. Users simply need to define the covariant metric tensor $g_{\mu\nu}$ of the target spacetime within a designated section of the Python script; \texttt{SpinOctopus} automatically computes the contravariant metric $g^{\mu\nu}$ and the required partial derivatives $g^{\mu\nu,r}$ and $g^{\mu\nu,\theta}$, subsequently generating the equation-of-motion module required for numerical integration.

Once the initial conditions of a particle, $\left(t,r,\theta,\varphi,p_{t},p_{r},p_{\theta},p_{\varphi}\right)$, are specified, \texttt{SpinOctopus} integrates timelike orbits using a fixed-step, sixth-order Runge-Kutta method (RK6), and traces photon trajectories utilizing an adaptive-step, fifth- and sixth-order Runge-Kutta-Fehlberg integrator (RKF56). Building upon the standard RKF56 algorithm, we implement an additional step-size control mechanism scaled by the radial coordinate $r$ and the polar angle $\theta$ of the ray. Specifically, the initial integration step-size $\Delta h$ provided to the RKF56 routine is continuously adjusted by monitoring the ray's position:
\begin{equation}\label{17}
\Delta h = h_{\textrm{min}}\left(\frac{r}{r_{\textrm{eh}}}\right)^{N}\frac{\sin\theta}{\sqrt{\sin^{2}\theta+\xi_{\theta}^{2}}}.
\end{equation}
Here, $h_{\textrm{min}}$ is the minimum base integration step-size preset in the code, typically on the order of $10^{-4}$---$10^{-3}$, while $N$ is the step-size amplification exponent, generally chosen in the range of $(1,2]$. When a ray enters the region where $\sin\theta < \xi_{\theta}$ (typically near the poles), the parameter $\xi_{\theta}$ effectively constrains the step-size, ensuring the ray traverses the coordinate singularity safely. In practice, $\xi_{\theta}$ is set within $[0.005, 0.05]$. Importantly, the specific combination of the parameters $h_{\textrm{min}}$, $N$, and $\xi_{\theta}$ profoundly dictates the balance between integration accuracy and computational efficiency; a detailed discussion on this is deferred to the subsequent section evaluating the code's performance.
\subsection{Initial conditions of light rays}
\begin{figure*}
\centering                   
\includegraphics[width=6cm]{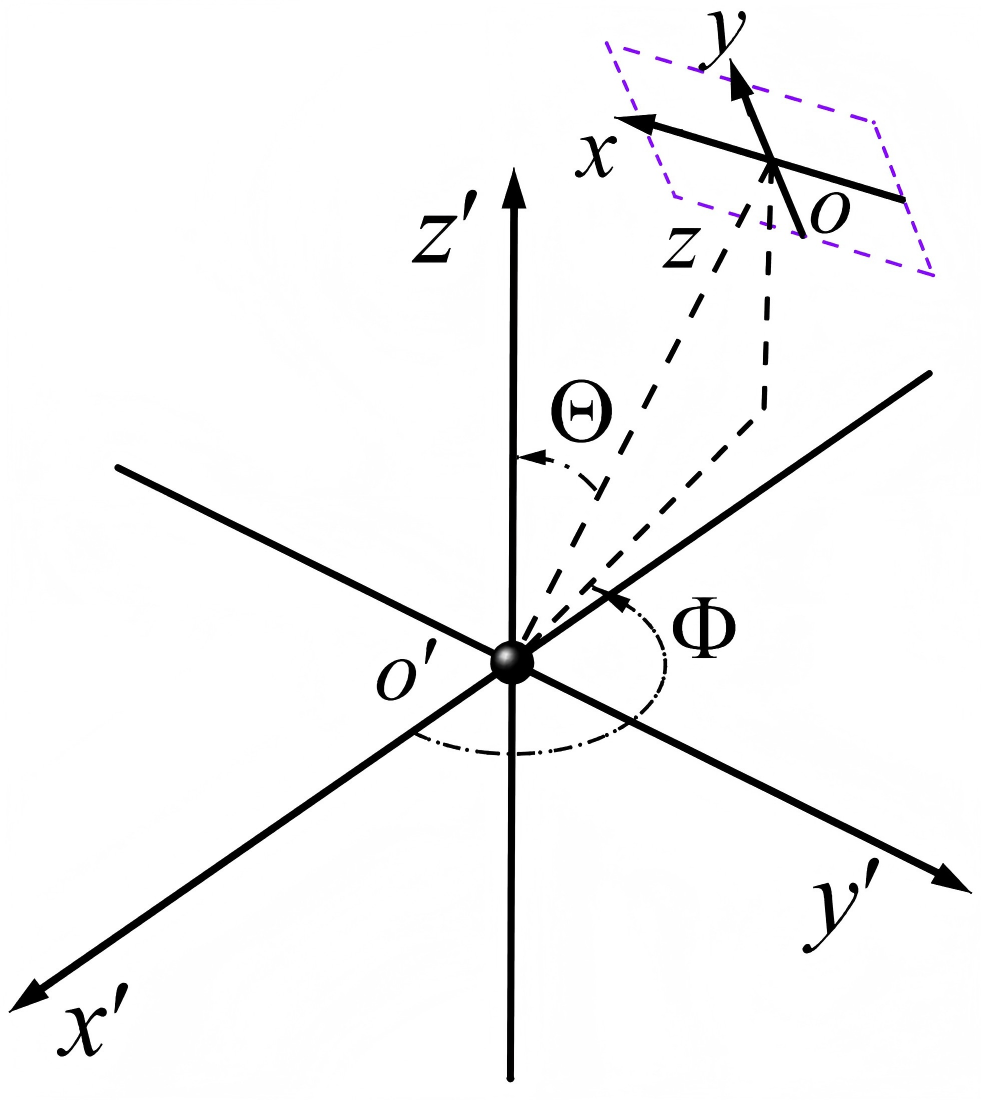}
\caption{Schematic illustration of the observer's local frame and the local frame of the curved spacetime adopted in the ray-tracing framework. The observer's space is characterized by the Cartesian coordinate system $xyz$, where the region enclosed by the purple dashed lines defines the observation screen. The compact object resides at $o^{\prime}$, which serves as the origin for the local coordinate system $x^{\prime}y^{\prime}z^{\prime}$. We define the angle between the rotation axis of the compact object, $o^{\prime}z^{\prime}$, and $o^{\prime}o$ as the observation inclination angle $\Theta$, and the angle between the projection of $o^{\prime}o$ onto the equatorial plane $x^{\prime}o^{\prime}y^{\prime}$ and the $x^{\prime}$-axis as the observation azimuthal angle $\Phi$.}      
\label{fig1}                 
\end{figure*}
The fundamental objective of ray-tracing is to establish a bridge connecting the observer and the light source residing in curved spacetime. In the physical picture, light rays are emitted by radiating matter within the curved spacetime and, following propagation, may be captured by the observer. Constructing a ray-tracing algorithm based on this forward-tracking approach would drastically inflate computational costs, given that only a fraction of the emitted rays successfully reach the observer. Consequently, practical implementations start from the observation screen and trace the detected rays backward in time, thereby establishing a backward ray-tracing framework. In this paradigm, the initial conditions of a ray are defined in the observer's local frame, whereas its trajectory is simulated within the curved spacetime (i.e., the local frame of the black hole). Therefore, a crucial preliminary step in backward ray-tracing is transforming the ray's initial conditions from the observer's frame into the local coordinate frame of the curved spacetime.

As illustrated in figure \ref{fig1}, the observer's local frame and the curved spacetime's local frame are denoted by $xyz$ and $x^{\prime}y^{\prime}z^{\prime}$, respectively. The purple dashed rectangle lying in the $xoy$ plane designates the observation screen. The $z$-axis of the observer's frame points directly toward the center of the compact object, which coincides with the origin $o^{\prime}$ of the $x^{\prime}y^{\prime}z^{\prime}$ system. The observation inclination angle $\Theta$ is defined as the angle between the line of sight $o^{\prime}o$ and the rotation axis $o^{\prime}z^{\prime}$. Similarly, the observation azimuthal angle $\Phi$ is the angle between the $x^{\prime}$-axis and the projection of $o^{\prime}o$ onto the equatorial plane $x^{\prime}o^{\prime}y^{\prime}$. Typically, the observer is placed at a sufficiently large distance; in our setup, the observation distance in the curved spacetime local frame is set to $r_{\textrm{obs}}=\overline{oo^{\prime}}=1000$.

For an arbitrary pixel chosen on the observation screen $xoy$, its coordinates $\left(x,y,z=0\right)$ denote the initial position of the ray from the observer's perspective. By rotating and translating the observer's local frame, $xyz$ can be aligned with $x^{\prime}y^{\prime}z^{\prime}$. This geometric operation yields the mapping relations that transform the coordinates $\left(x,y,0\right)$ into the curved spacetime's local frame $x^{\prime}y^{\prime}z^{\prime}$ \cite{Pu:2016eml,Younsi:2016azx}:
\begin{eqnarray}
x^{\prime} &=& \mathcal{T}\cos\Phi - x\sin\Phi, \label{18} \\
y^{\prime} &=& \mathcal{T}\sin\Phi + x\cos\Phi, \label{19} \\
z^{\prime} &=& \left(r_{\textrm{obs}} - z\right)\cos\Theta + y\sin\Theta, \label{20}
\end{eqnarray}
where the function $\mathcal{T}$ takes the form
\begin{equation}\label{21}
\mathcal{T} = \left(\sqrt{r_{\textrm{obs}}^{2}+a^{2}}-z\right)\sin\Theta-y\cos\Theta,
\end{equation}
with $a$ denoting the dimensionless spin parameter. Subsequently, the spatial coordinates in equations \eqref{18}---\eqref{20} can be transformed into the Boyer-Lindquist coordinates $x^{\mu}=\left(t,r,\theta,\varphi\right)$. By setting the initial coordinate time of the light ray to $t=0$, we obtain
\begin{eqnarray}
t &=& 0, \label{22} \\
r &=& \sqrt{\frac{\mathcal{D}+\sqrt{\mathcal{D}^{2}+4a^{2}z^{\prime2}}}{2}}, \label{23} \\
\theta &=& \arccos\left(\frac{z^{\prime}}{r}\right), \label{24} \\
\varphi &=& \textrm{atan}2\left(y^{\prime},x^{\prime}\right), \label{25}
\end{eqnarray}
where $\mathcal{D}$ is given by
\begin{equation}\label{26}
\mathcal{D} = x^{\prime2} + y^{\prime2} + z^{\prime2} - a^{2}.
\end{equation}

Given the large observation distance, the spacetime at the observer's location can be approximated as flat. Consequently, from a forward-tracking perspective, rays strike the observation screen along paths nearly parallel to the line of sight $o^{\prime}o$. Thus, the initial spatial velocity of a ray can be naturally set to $\left(\dot{x},\dot{y},\dot{z}\right)=\left(0,0,-1\right)$. Differentiating the coordinate transformation relations in equations \eqref{18}---\eqref{20} and substituting the velocity vector $\left(0,0,-1\right)$, we obtain
\begin{eqnarray}
\dot{x}^{\prime} &=& \sin\Theta\cos\Phi, \label{27} \\
\dot{y}^{\prime} &=& \sin\Theta\sin\Phi, \label{28} \\
\dot{z}^{\prime} &=& \cos\Theta. \label{29}
\end{eqnarray}
Furthermore, substituting $\left(\dot{x}^{\prime},\dot{y}^{\prime},\dot{z}^{\prime}\right)$ into the differential forms of equations \eqref{23}---\eqref{25} yields
\begin{eqnarray}
\dot{r} &=& \frac{r\mathcal{R}\sin\theta\sin\Theta\cos\Psi+\mathcal{R}^{2}\cos\theta\cos\Theta}{r^{2}+a^{2}\cos^{2}\theta}, \label{30} \\
\dot{\theta} &=& \frac{\mathcal{R}\cos\theta\sin\Theta\cos\Psi-r\sin\theta\cos\Theta}{r^{2}+a^{2}\cos^{2}\theta}, \label{31} \\
\dot{\varphi} &=& -\frac{\sin\Theta\sin\Psi}{\mathcal{R}\sin\theta}, \label{32}
\end{eqnarray}
where $\mathcal{R}=\sqrt{r^{2}+a^{2}}$ and $\Psi=\varphi-\Phi$.

Next, we need to determine the initial conjugate momenta of the light ray. Specifically, $p_{r}$ and $p_{\theta}$ can be readily evaluated using equations \eqref{5} and \eqref{6}, whereas determining $p_{t}$ and $p_{\varphi}$ requires solving for $\dot{t}$ first. Substituting equations \eqref{30}---\eqref{32} into the Lagrangian \eqref{3} and imposing the null constraint, we solve for $\dot{t}$ as
\begin{equation}\label{33}
\dot{t} = \frac{-f_{2} + \sqrt{f_{2}^{2}-4f_{1}f_{3}}}{2f_{1}},
\end{equation}
where $f_{1}=g_{tt}$, $f_{2}=2g_{t\varphi}\dot{\varphi}$, and $f_{3}=g_{rr}\dot{r}^{2}+g_{\theta\theta}\dot{\theta}^{2}+g_{\varphi\varphi}\dot{\varphi}^{2}$. Substituting the expressions for $\dot{t}$ and $\dot{\varphi}$ from equations \eqref{33} and \eqref{32} into equations \eqref{4} and \eqref{7} yields $p_{t}$ and $p_{\varphi}$.

Because the propagation path of a null geodesic is independent of the photon frequency, the spatial momentum components can be renormalized by $p_{t}$ \cite{Cunha:2016bpi}, yielding $\hat{p}_{r}=p_{r}/p_{t}$, $\hat{p}_{\theta}=p_{\theta}/p_{t}$, and $\hat{p}_{\varphi}=p_{\varphi}/p_{t}$. Consequently, the initial canonical variables of the photon, designated for tracing the ray within the curved spacetime's local frame, are set to $\left(0,r,\theta,\varphi,-1,\hat{p}_{r},\hat{p}_{\theta},\hat{p}_{\varphi}\right)$. Integrating the system of differential equations \eqref{9}---\eqref{16} using the aforementioned RKF56 algorithm based on these initial conditions ultimately yields the propagation path of the ray. It is crucial to emphasize that because \texttt{SpinOctopus} operates on a backward ray-tracing paradigm, the integration must be performed using a negative step-size; alternatively, one may use a positive step-size if an overall minus sign is introduced into the differential equations.
\subsection{Redshift factor and radiative transfer}
In the previous subsection, we addressed the propagation of rays within curved spacetime. Physically, rays traced backward from the observation screen encounter one of three possible fates. First, they may escape to infinity without interacting with any radiating medium along their propagation paths. Second, they may plunge directly into the black hole event horizon without encountering radiating matter; the pixels corresponding to these rays constitute the black hole shadow. Third, they may intersect with radiating matter during propagation, in which case they will carry specific intensity. According to the covariant radiative transfer equation, as a ray traverses a medium, the specific intensity it carries satisfies
\begin{equation}\label{34}
\frac{\textrm{d}\mathcal{I}}{\textrm{d}\lambda} = \mathcal{J} - \mathcal{A}\mathcal{I},
\end{equation}
where $\mathcal{I}=I_{\nu}/\nu^{3}$, $\mathcal{A}=\nu\alpha_{\nu}$, and $\mathcal{J}=j_{\nu}/\nu^{2}$ are the invariants of the specific intensity, absorption coefficient, and emission coefficient, respectively, with $\nu$ denoting the emission frequency. Subsequently, the above equation can be expanded as
\begin{equation}\label{35}
\frac{\textrm{d}}{\textrm{d}\lambda}\left(\frac{I_{\nu}}{\nu^{3}}\right) = \frac{j_{\nu}}{\nu^{2}}-\nu\alpha_{\nu}\left(\frac{I_{\nu}}{\nu^{3}}\right).
\end{equation}

We define the redshift factor $g$ as
\begin{equation}\label{36}
g = \frac{\nu_{\textrm{obs}}}{\nu}=\frac{\left(p_{\mu}\dot{x}^{\mu}\right)_{\textrm{obs}}}{\left(p_{\mu}\dot{x}^{\mu}\right)_{\textrm{source}}},
\end{equation}
where $\dot{x}^{\mu}$ denotes the four-velocity of either the observer or the radiation source, corresponding to the subscript ``\textrm{obs}'' or ``\textrm{source}'', respectively. For a stationary observer, this four-velocity can be fixed to $\dot{x}_{\textrm{obs}}^{\mu}=(1,0,0,0)$. Substituting the redshift factor into equation \eqref{35} yields
\begin{equation}\label{37}
\frac{\textrm{d}}{\textrm{d}\lambda}\left(g^{3}\frac{I_{\nu}}{\nu_{\textrm{obs}}^{3}}\right) = g^{2}\frac{j_{\nu}}{\nu_{\textrm{obs}}^{2}} - \alpha_{\nu}\frac{\nu_{\textrm{obs}}}{g}\left(g^{3}\frac{I_{\nu}}{\nu_{\textrm{obs}}^{3}}\right).
\end{equation}

It is worth noting that, from a phenomenological perspective, by adopting frequency-independent emission $j_{\nu}$ and absorption $\alpha_{\nu}$ in the local frame and fixing $\nu_{\textrm{obs}}=1$, all frequency shift effects on the specific intensity of the ray are encapsulated entirely within the redshift factor. Consequently, equation \eqref{37} is rewritten as
\begin{equation}\label{38}
\frac{\textrm{d}I_{\textrm{obs}}}{\textrm{d}\lambda} = g^{2}j_{\nu} - \frac{\alpha_{\nu}}{g}I_{\textrm{obs}},
\end{equation}
where $I_{\textrm{obs}}$ denotes the specific intensity of the ray corresponding to a given pixel on the observation screen.

During the ray-tracing procedure, particularly as the ray navigates through the radiating medium near the central compact object, the orbital integration step-size is extremely small. Under this small-step-size approximation, equation \eqref{38} can be split into two differential equations:
\begin{eqnarray}
\frac{\textrm{d}\varrho}{\textrm{d}\lambda} &=& \frac{\alpha_{\nu}}{g}, \label{39} \\
\frac{\textrm{d}I_{\textrm{obs}}}{\textrm{d}\lambda} &=& g^{2}j_{\nu}\textrm{e}^{-\varrho}, \label{40} 
\end{eqnarray}
where $\varrho$ denotes the optical depth. Once the geometric structure of the radiation source is specified, the ray's travel path within the source can be determined. At this stage, integrating the differential equations \eqref{39} and \eqref{40} simultaneously with the photon equations of motion \eqref{9}---\eqref{16} yields the contribution of this trajectory segment to the specific intensity of the ray.
\section{Functionalities and Efficiency of the Code}
In this section, taking the Kerr black hole as a representative example, we demonstrate the functionalities of \texttt{SpinOctopus}. These include simulating black hole critical curves and images under various light sources, solving for timelike and null orbits, and simulating light curves for dynamic radiating sources.
\subsection{Kerr metric}
In Boyer-Lindquist coordinates $x^{\mu}=\left(t,r,\theta,\varphi\right)$, the dimensionless covariant metric components of the Kerr black hole are written as \cite{Cunha:2016bpi,Wu:2021rrd}
\begin{eqnarray}
g_{tt} &=& -\left(1-\frac{2r}{r^{2}+a^{2}\cos^{2}\theta}\right), \label{41} \\
g_{rr} &=& \frac{r^{2}+a^{2}\cos^{2}\theta}{r^{2}-2r+a^{2}}, \label{42} \\
g_{\theta\theta} &=& r^{2}+a^{2}\cos^{2}\theta, \label{43} \\
g_{\varphi\varphi} &=& \left(r^{2}+a^{2}+\frac{2a^{2}r\sin^{2}\theta}{r^{2}+a^{2}\cos^{2}\theta}\right)\sin^{2}\theta, \label{44} \\
g_{t\varphi} &=& -\frac{2ar\sin^{2}\theta}{r^{2}+a^{2}\cos^{2}\theta}. \label{45}
\end{eqnarray}
In subsequent calculations, we fix the spin parameter to $a=0.94$. By solving $\Delta\left(r\right)=r^{2}-2r+a^{2}=0$, we find that the event horizon radius of the Kerr black hole is fixed at $r_{\textrm{eh}} \approx 1.341$.
\subsection{Photon propagation}
\begin{figure*}
\centering                   
\includegraphics[width=14cm]{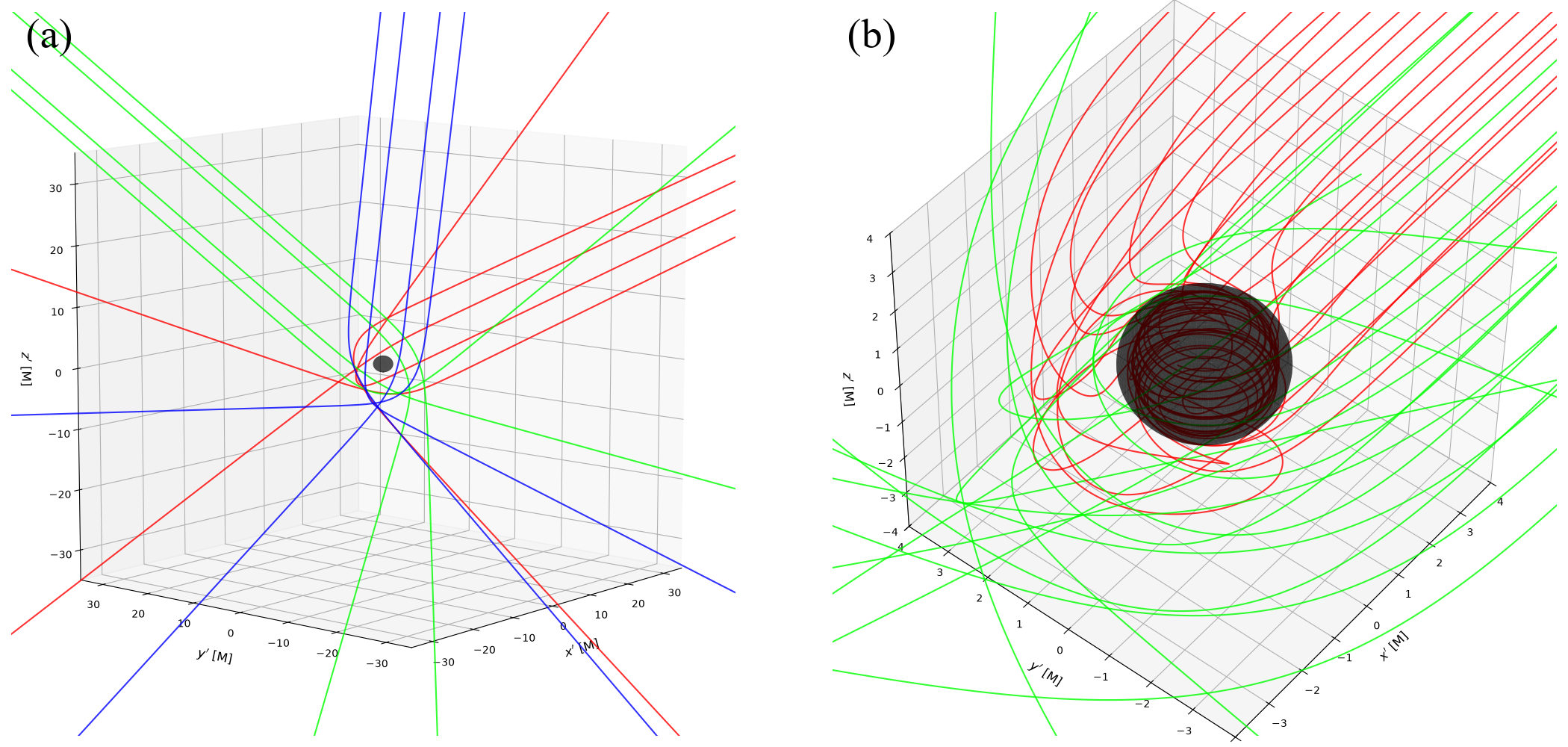}
\caption{Schematic diagram of ray propagation in the Kerr spacetime across different parameter spaces. In both panels, the black transparent ellipsoid represents the event horizon of the Kerr black hole. In panel (a), the red, green, and blue rays correspond to the observation parameters $\left(\Theta,\Phi\right)$ of $\left(80^{\circ},0^{\circ}\right)$, $\left(50^{\circ},120^{\circ}\right)$, and $\left(17^{\circ},240^{\circ}\right)$, respectively. Originating from different grid points on the observation screen, the rays are deflected by the black hole and all escape to infinity. In panel (b), the red and green rays originate from the observation screen corresponding to $\left(\Theta,\Phi\right)=\left(90^{\circ},0^{\circ}\right)$, subsequently plunging into the black hole and escaping to infinity, respectively. All ray simulations adopt the same integration parameters: $h_{\textrm{min}}=0.0005$, $N=1.8$, $\xi_{\theta}=0.01$, and an RKF56 tolerance of $10^{-8}$.}      
\label{fig2}                 
\end{figure*}
By arbitrarily selecting a pixel coordinate $\left(x,y\right)$ on the observation screen, we can trace the trajectory of the ray in the curved spacetime, denoted as $\left(t,r,\theta,\varphi\right)$. By transforming the Boyer-Lindquist coordinates into oblate spheroidal coordinates,
\begin{eqnarray}
x^{\prime} &=& \sqrt{r^{2}+a^{2}}\sin\theta\cos\varphi, \label{81} \\
y^{\prime} &=& \sqrt{r^{2}+a^{2}}\sin\theta\sin\varphi, \label{82} \\
z^{\prime} &=& r\cos\theta, \label{83}
\end{eqnarray}
we obtain the propagation trajectory of the ray in three-dimensional space.

We select three different observer positions, with observation inclination and azimuthal angles $\left(\Theta,\Phi\right)$ set to $\left(80^{\circ},0^{\circ}\right)$, $\left(50^{\circ},120^{\circ}\right)$, and $\left(17^{\circ},240^{\circ}\right)$, respectively. For each observation screen, four pixel points are chosen: $\left(x,y\right)=\left(5,5\right)$, $\left(5,-5\right)$, $\left(-5,-5\right)$, and $\left(-5,5\right)$, serving as the starting points for backward ray-tracing. These 12 ray trajectories are illustrated in figure \ref{fig2}(a). Each ray propagates to the vicinity of the black hole, undergoes deflection, and eventually escapes to infinity (to halt the integration, an artificial cutoff is imposed when the ray's radial coordinate satisfies $r \geq 1200$). The number of integration steps for a single ray falls within the interval $[3100,3800]$.

Fixing the observation parameters to $\left(\Theta,\Phi\right)=\left(90^{\circ},0^{\circ}\right)$, we uniformly sample the observation screen within $x,y \in [-5,5]$ to trace 36 rays, as shown in figure \ref{fig2}(b). Among them, the red rays plunge into the black hole event horizon,\footnote{\texttt{SpinOctopus} determines whether a ray falls into the black hole by monitoring if its radial coordinate satisfies $r \leq r_{\textrm{eh}}+10^{-3}$. Although $10^{-3}$ is effective in most cases, this threshold may need to be reduced when dealing with the $\gamma$ metric \cite{2019PhRvD.100b4014A,2022EPJC...82..696S}.} whereas the green rays escape to infinity. The frame-dragging effect induced by the black hole spin is also clearly discernible. Furthermore, it is observed that the escaping rays predominantly originate from the region where $y < 0$. This occurs because the combination of parameters $a=0.94$ and $\Theta=90^{\circ}$ causes the observed black hole shadow to assume a classic ``D'' shape; the deformation of the shadow on its left side effectively creates a window for the escaping rays.
\subsection{Shadow profiles}
\begin{figure*}
\centering                   
\includegraphics[width=8cm]{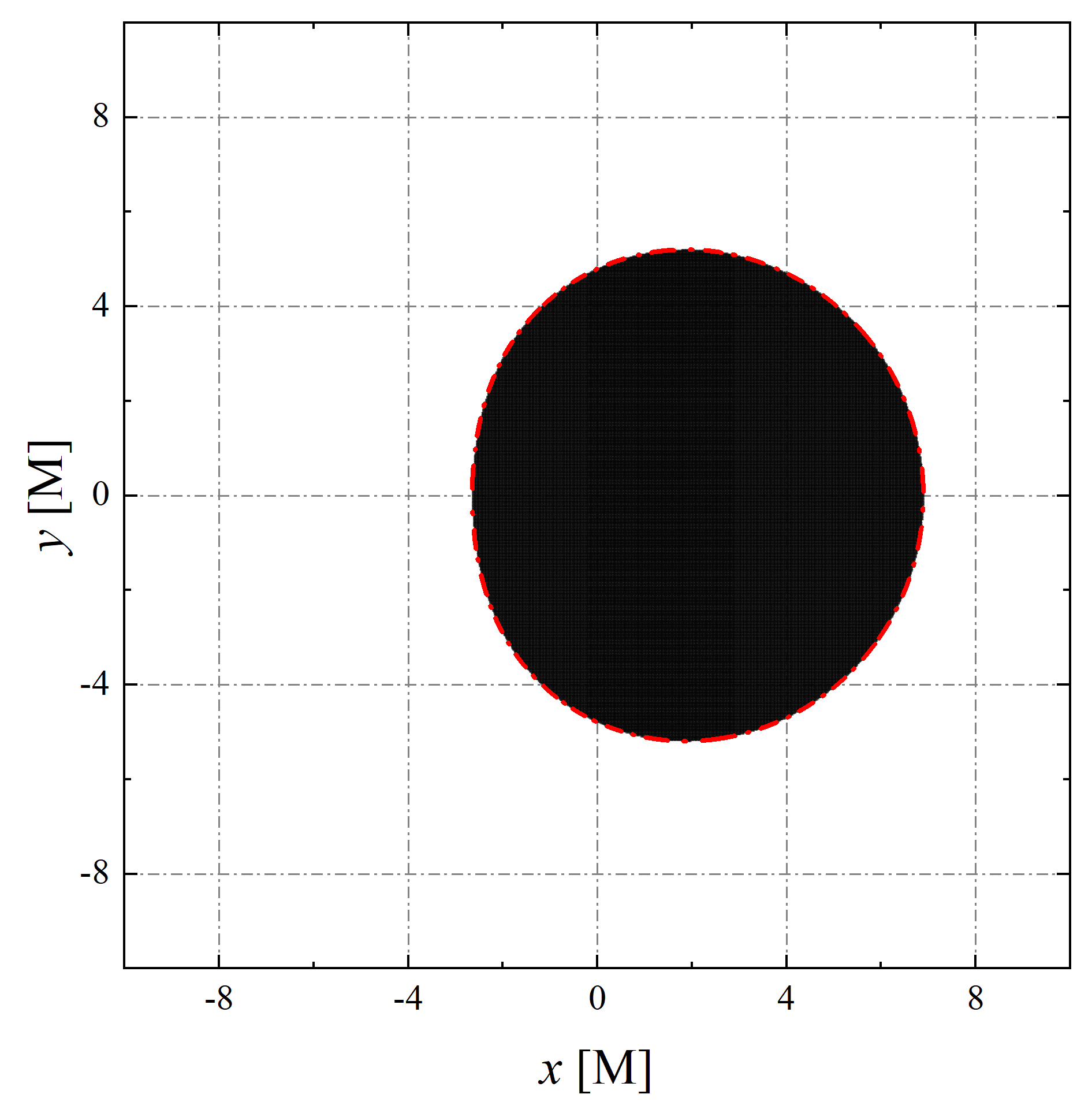}
\caption{The shadow of a Kerr black hole at an observation inclination of $\Theta=90^{\circ}$. The black region represents the \texttt{SpinOctopus} simulation result, while the red dashed line shows the analytical calculation. The excellent agreement demonstrates our code's high precision in simulating black hole shadows. Here, the adaptive step-size parameters are set to $h_{\textrm{min}}=0.001$, $N=1.6$, and $\xi_{\theta}=0.01$, with an RKF56 tolerance of $10^{-8}$.}      
\label{fig3}                 
\end{figure*}

We set the field of view to $x,y \in [-10,10]$ with a resolution of 1000 $\times$ 1000 pixels. Figure \ref{fig3} shows the shadow of a Kerr black hole at an observation angle of $\Theta=90^{\circ}$. The black region represents the simulation result obtained by \texttt{SpinOctopus}, while the red dashed line corresponds to the analytical shadow profile, the calculation method for which is widely available in the literature \cite{2022PhR...947....1P}. Visually, the red dashed line aligns precisely with the edge of the black region. To quantify this, we extract the maximum and minimum horizontal and vertical coordinates of the black pixel set, denoted as $x_{\textrm{max}}^{\textrm{S}}=6.8969$, $x_{\textrm{min}}^{\textrm{S}}=-2.6326$, $y_{\textrm{max}}^{\textrm{S}}=5.1952$, and $y_{\textrm{min}}^{\textrm{S}}=-5.1952$, respectively. Using the analytical formulas, we calculate the corresponding theoretical values: $x_{\textrm{max}}^{\textrm{A}}=6.8996$, $x_{\textrm{min}}^{\textrm{A}}=-2.6415$, $y_{\textrm{max}}^{\textrm{A}}=5.1962$, and $y_{\textrm{min}}^{\textrm{A}}=-5.1962$. Comparing the analytical solutions with the \texttt{SpinOctopus} results at these four key positions reveals a maximum deviation of $\varepsilon = |x_{\textrm{min}}^{\textrm{S}}-x_{\textrm{min}}^{\textrm{A}}| = 8.9 \times 10^{-3}$. Taking Sgr A$^{*}$ as an example, its angular scale is $GM/\left(c^{2}D\right) \approx 5.03$ $\mu$as [], where $D$ is the distance from the black hole to Earth. The angular discrepancy introduced by $\varepsilon$ is therefore only $0.045$ $\mu$as, which is far below the current angular resolution of the EHT. This confirms that \texttt{SpinOctopus} achieves extremely high accuracy in simulating black hole shadows (critical curves).
\subsection{Imaging in a multi-component accretion environment}
\subsubsection{Emission and absorption}
In our previous work, we developed an analytical model for a multi-component accretion environment and focused on the imaging features of a Kerr black hole against this background \cite{Hu:2026zpn}. Here, the multi-component accretion environment serves as one of the built-in light sources in \texttt{SpinOctopus}. Its emission function is composed of multiple contributions, including a geometrically thick accretion disk $j_{\textrm{d}}\left(r,\theta\right)$, a compact radiating bright ring within the disk characterized by a Gaussian bump $j_{\textrm{b}}\left(r,\theta\right)$, and a local compact radiating hotspot also described by a Gaussian profile $j_{\textrm{s}}\left(r,\theta,\varphi\right)$. Consequently, the total emission $j_{\nu}\left(r,\theta,\varphi\right)$ of the accretion environment in the local frame is formulated as
\begin{equation}\label{46}
j_{\nu}\left(r,\theta,\varphi\right)=j_{0}\left[j_{1}j_{\textrm{d}}\left(r,\theta\right)+j_{2}j_{\textrm{b}}\left(r,\theta\right)+j_{3}j_{\textrm{s}}\left(r,\theta,\varphi\right)\right].
\end{equation}
Here, $j_{i}$ ($i=1,2,3$) dictates the emission weight of each respective component, while $j_{0}$ determines the overall emission intensity of the entire accretion environment. For simplicity in subsequent calculations, these parameters are all fixed to unity; if a specific component needs to be omitted, its corresponding weight is set to $j_{i}=0$.

We assume that the structure of the geometrically thick accretion disk is symmetric about the equatorial plane. Its emission originates from the inner boundary $r_{\textrm{in}}$ and undergoes radial decay, while a Gaussian profile is employed to broaden it in the vertical space. Thus, in logarithmic space, we have
\begin{equation}\label{47}
\ln\left[j_{\textrm{d}}\left(r,\theta\right)\right]=p_{1}\ln\left(\frac{r_{\textrm{eff}}}{r_{\textrm{in}}}\right)+p_{2}\left[\ln\left(\frac{r_{\textrm{eff}}}{r_{\textrm{in}}}\right)\right]^{2}-\frac{\left(\theta-\pi/2\right)^{2}}{2\left(\sigma_{d\theta} \textrm{e}^{r\delta_{d}}\right)^{2}},
\end{equation}
where $p_{1}$ and $p_{2}$ finely govern the radial decay of the accretion disk; $\sigma_{d\theta}$ determines the initial thickness at the disk's inner boundary, and $\delta_{d}$ dictates the expansion rate of the disk in the vertical direction. Depending on the combination of $\sigma_{d\theta}$ and $\delta_{d}$, the vertical profile in the $r\sin\theta$ versus $r\cos\theta$ plane can exhibit a rich variety of geometric structures, such as conical, trapezoidal, and trumpet-like shapes. Additionally, by introducing the effective radius $r_{\textrm{eff}}$, we establish a radiation compensation region for the accretion disk. Within this region, the radial decay is suppressed and is instead characterized by a plateau or a bump. The expression for $r_{\textrm{eff}}$ is formulated as
\begin{equation}\label{48}
r_{\textrm{eff}}=r-j_{\textrm{p}}w_{\textrm{p}}\left[\tanh(\frac{r-r_{\textrm{p}}}{w_{\textrm{p}}})-\tanh(\frac{r_{\textrm{in}}-r_{\textrm{p}}}{w_{\textrm{p}}})\right],
\end{equation}
where $j_{\textrm{p}}$ controls the emission intensity of the radiation compensation region. Specifically, when $j_{\textrm{p}}=1$, the radial profile of this compensation region manifests as a plateau; when $j_{\textrm{p}} > 1$, it takes on a raised shape; and when $j_{\textrm{p}} < 1$, the profile exhibits a downward slope. In particular, the radiation compensation can be entirely deactivated by setting $j_{\textrm{p}}=0$. The remaining parameters, $w_{\textrm{p}}$ and $r_{\textrm{p}}$, denote the radial width and the central position of the compensation region, respectively.

Geometrically thick accretion disks often host a rich variety of high-energy astrophysical events, such as outward-propagating shock waves, which spatially resemble a compact radiating torus. Although a shock is an exceptionally complex magnetohydrodynamic structure, from a phenomenological perspective, we can qualitatively characterize it using a Gaussian distribution:
\begin{equation}\label{49}
\ln\left[j_{\textrm{b}}\left(r,\theta\right)\right]=-\frac{\left(r-r_{\textrm{b}}\right)^{2}}{2\sigma_{br}^{2}}-\frac{\left(\theta-\pi/2\right)^{2}}{2\sigma_{b\theta}^{2}}.
\end{equation}
Here, $r_{\textrm{b}}$ designates the radial position of the shock, while $\sigma_{br}$ and $\sigma_{b\theta}$ govern its geometric morphology. According to our previous study \cite{Hu:2026zpn}, in the vertical cross-section, the shock can manifest as bow-like, quasi-circular, or teardrop-like shapes, depending on the specific combination of $\sigma_{br}$ and $\sigma_{b\theta}$. Furthermore, the introduction of the shock is equivalent to superimposing a bump onto the geometrically thick accretion disk background; therefore, we also refer to this component as a Gaussian bump.

Analogous to the shock wave $j_{\textrm{b}}\left(r,\theta\right)$, we introduce $j_{\textrm{s}}\left(r,\theta,\varphi\right)$ to qualitatively characterize localized compact radiating hotspots induced by high-energy phenomena within the disk, such as matter collisions, radiation from a companion star, tidal disruption events (TDE), and magnetic reconnection events. In logarithmic space, it is expressed as
\begin{equation}\label{50}
\ln\left[j_{\textrm{s}}\left(r,\theta,\varphi\right)\right]=-\frac{\left(r-r_{\textrm{s}}\right)^{2}}{2\sigma_{sr}^{2}}-\frac{\left(\theta-\theta_{\textrm{s}}\right)^{2}}{2\sigma_{s\theta}^{2}}-\frac{\left(\varphi-\varphi_{\textrm{s}}\right)^{2}}{2\sigma_{s\varphi}^{2}},
\end{equation}
where $\left(r_{\textrm{s}},\theta_{\textrm{s}},\varphi_{\textrm{s}}\right)$ specifies the spatial coordinates of the hotspot, and $\sigma_{s\mu}$ (with the index $\mu \in \{r,\theta,\varphi\}$) dictates its geometric morphology. It is worth noting that a localized hotspot may undergo tidal deformation under the influence of the strong gravitational field, acquiring a pear-like shape with its apex pointing toward the central compact object; the emission function \eqref{50} naturally captures this characteristic.

To minimize model complexity, the absorption function adopts an identical functional form to $j_{\nu}\left(r,\theta,\varphi\right)$:\footnote{Unless otherwise specified in subsequent light source models, we consistently adopt this convention, assigning identical analytical forms to both the emission and absorption functions.}
\begin{eqnarray}\label{51}
\alpha_{\nu}\left(r,\theta,\varphi\right)=\alpha_{0}\left[\alpha_{1}\alpha_{\textrm{d}}\left(r,\theta\right)+\alpha_{2}\alpha_{\textrm{b}}\left(r,\theta\right)+\alpha_{3}\alpha_{\textrm{s}}\left(r,\theta,\varphi\right)\right].
\end{eqnarray}
Here, $\alpha_{\textrm{d}}\left(r,\theta\right)$, $\alpha_{\textrm{b}}\left(r,\theta\right)$, and $\alpha_{\textrm{s}}\left(r,\theta,\varphi\right)$ represent the absorption functions of the geometrically thick accretion disk, the Gaussian bump, and the localized hotspot, respectively. The absorption intensities of the various components are solely governed by adjusting the weight factors $\alpha_{i}$ (for $i=0,1,2,3$). Setting $\alpha_{i}=0$ implies that the corresponding component is optically thin.
\subsubsection{Dynamics}
At present, the majority of studies partition the motion domain of accreting matter into a plunging region and a Keplerian region, separated by the innermost stable circular orbit (ISCO) of timelike particles. Furthermore, the dynamical solutions for the matter typically rely on constants of motion, such as specific energy, specific angular momentum, and the Carter constant. Although highly effective, this approach lacks sufficient degrees of freedom and exhibits limited universality. To address this, \texttt{SpinOctopus} adopts a global inspiral approximation to prescribe the dynamics of the accreting matter. Specifically, initiating from an artificially prescribed outer boundary $r_{\textrm{out}}$ of the accretion environment, the accreting matter accelerates as it inspirals toward the central compact object, simultaneously converging toward the equatorial plane.

In the local frame, the physical three-velocity $(\hat{v}^{r},\hat{v}^{\theta},\hat{v}^{\varphi})$ of the accreting matter is defined as
\begin{equation}\label{52}
\hat{v}^{r}\left(r\right) = -V_{\textrm{max}}^{r}\left(\frac{r_{\textrm{in}}}{r}\right)^{p_{3}},
\end{equation}
\begin{equation}\label{53}
\hat{v}^{\theta}\left(r,\theta\right) = -V_{\textrm{max}}^{\theta}\left(\theta-\frac{\pi}{2}\right)\left(\frac{r_{\textrm{in}}}{r}\right)^{p_{4}},
\end{equation}
and
\begin{equation}\label{54}
\hat{v}^{\varphi}\left(r\right) = V_{\textrm{max}}^{\varphi}\left(\frac{r_{\textrm{in}}}{r}\right)^{p_{5}}.
\end{equation}
Here, $p_{3}$, $p_{4}$, and $p_{5}$ measure the rates of variation for the velocity components in the three respective directions. The parameter $V_{\textrm{max}}^{\mu}$ designates the upper limit for each velocity component and simultaneously provides a parameter window to truncate the velocity in a specific direction; for instance, setting $V_{\textrm{max}}^{\theta}=V_{\textrm{max}}^{\varphi}=0$ results in the matter radially plunging into the black hole. Furthermore, the negative signs in the expressions for $\hat{v}^{r}\left(r\right)$ and $\hat{v}^{\theta}\left(r,\theta\right)$ indicate that the matter moves inward toward the central object and that the equatorial plane acts as an attractor, respectively.

Subsequently, the physical velocity of the accreting matter must be transformed into the coordinate velocity to facilitate the calculation of the redshift factor $g$ \eqref{36}. To achieve this, we introduce a Zero Angular Momentum Observer (ZAMO), whose tetrad basis is given by \cite{Cunha:2016bpi,Wang:2021ara,Hou:2022eev}
\begin{eqnarray}
\hat{e}_{t} &=& \frac{1}{\kappa}\left(\partial_{t}+\Omega\partial_{\varphi}\right), \label{55} \\
\hat{e}_{r} &=& \frac{1}{\sqrt{g_{rr}}}\partial_{r}, \label{56} \\
\hat{e}_{\theta} &=& \frac{1}{\sqrt{g_{\theta\theta}}}\partial_{\theta}, \label{57} \\
\hat{e}_{\varphi} &=& \frac{1}{\sqrt{g_{\varphi\varphi}}}\partial_{\varphi}, \label{58}
\end{eqnarray}
where
\begin{eqnarray}
\kappa = \sqrt{-g_{tt}+\frac{g_{t\varphi}^{2}}{g_{\varphi\varphi}}} \label{59}
\end{eqnarray}
and 
\begin{equation}\label{60}
\Omega = -\frac{g_{t\varphi}}{g_{\varphi\varphi}}
\end{equation}
denote the lapse function and the angular velocity of the frame-dragging effect, respectively. Next, the Lorentz factor $\gamma$ is defined as
\begin{equation}\label{61}
\gamma = \frac{1}{\sqrt{1-\left(\hat{v}^{r}\right)^{2}-\left(\hat{v}^{\theta}\right)^{2}-\left(\hat{v}^{\varphi}\right)^{2}}}.
\end{equation}
Consequently, the four-velocity of the accreting matter, as measured by the ZAMO, can be obtained through the projection
\begin{equation}\label{62}
\dot{x}^{\mu} = \gamma\left(\hat{e}_{t}^{\mu}+\hat{v}^{\nu}\hat{e}_{\nu}^{\mu}\right),
\end{equation}
where the spatial index $\nu$ runs over $\{r,\theta,\varphi\}$. Written out explicitly, this yields
\begin{eqnarray}
\dot{x}^{t} &=& \dot{t} = \frac{\gamma}{\kappa}, \label{63} \\
\dot{x}^{r} &=& \dot{r} = \frac{\gamma\hat{v}^{r}}{\sqrt{g_{rr}}}, \label{64} \\
\dot{x}^{\theta} &=& \dot{\theta} = \frac{\gamma\hat{v}^{\theta}}{\sqrt{g_{\theta\theta}}}, \label{65} \\
\dot{x}^{\varphi} &=& \dot{\varphi} = \gamma\left(\frac{\Omega}{\kappa}+\frac{\hat{v}^{\varphi}}{\sqrt{g_{\varphi\varphi}}}\right). \label{66}
\end{eqnarray}
\subsubsection{Image features}
Setting the field of view to $x,y \in [-15,15]$ with a resolution of $1000 \times 1000$ pixels, we present the Kerr black hole images for three distinct scenarios across various observation inclination angles. The first scenario, denoted as Case$_{\textrm{A}}$, considers only the contributions from the geometrically thick accretion disk and the radiation compensation in the emission function \eqref{46}, setting $j_{2}=j_{3}=0$. The second scenario, denoted as Case$_{\textrm{B}}$, builds upon Case$_{\textrm{A}}$ by superimposing the contribution of the Gaussian bump ($j_{2}=1$) while maintaining $j_{3}=0$. The final scenario, denoted as Case$_{\textrm{C}}$, further incorporates the localized compact hotspot ($j_{3}=1$) into the configuration of Case$_{\textrm{B}}$. Furthermore, the absorption of rays by the accreting matter is neglected across all simulations.
\begin{table}[htbp]
    \centering
    \renewcommand{\arraystretch}{1.2} 
    \setlength{\arrayrulewidth}{0.8pt}
    \caption{Model parameters of the multi-component accretion environments for Case$_{\textrm{A}}$, Case$_{\textrm{B}}$, and Case$_{\textrm{C}}$.} 
    \label{tab1}
    \begin{tabular}{|>{\centering\arraybackslash}m{3cm}|>{\centering\arraybackslash}m{3cm}|>{\centering\arraybackslash}m{3cm}|>{\centering\arraybackslash}m{3cm}|}
        \hline
        Parameters & Case$_{\textrm{A}}$ & Case$_{\textrm{B}}$ & Case$_{\textrm{C}}$ \\
        \hline
        \(p_1\) & \multicolumn{3}{c|}{-1.5} \\
        \hline
        \(p_2\) & \multicolumn{3}{c|}{-0.5} \\
        \hline
        \(\sigma_{d\theta}\) & \multicolumn{3}{c|}{0.2} \\
        \hline
        \(\delta_d\) & \multicolumn{3}{c|}{0.1} \\
        \hline
        \(j_\textrm{p}\) & \multicolumn{3}{c|}{1.0} \\
        \hline
        \(r_\textrm{p}\) & \multicolumn{3}{c|}{5.0} \\
        \hline
        \(w_\textrm{p}\) & \multicolumn{3}{c|}{3.0} \\
        \hline
        \(r_\textrm{b}\) & \multirow{3}{*}{0.0} & \multicolumn{2}{c|}{9.0} \\
        \cline{1-1} \cline{3-4}
        \(\sigma_{br}\) & & \multicolumn{2}{c|}{1.0} \\
        \cline{1-1} \cline{3-4}
        \(\sigma_{b\theta}\) & & \multicolumn{2}{c|}{0.2} \\
        \hline
        \(\left(r_\textrm{s},\theta_\textrm{s},\varphi_\textrm{s}\right)\) & \multicolumn{2}{c|}{\multirow{4}{*}{0.0}} & $\left(11,90^{\circ},160^{\circ}\right)$ \\
        \cline{1-1} \cline{4-4}
        \(\sigma_{sr}\) & \multicolumn{2}{c|}{} & 2.0 \\
        \cline{1-1} \cline{4-4}
        \(\sigma_{s\theta}\) & \multicolumn{2}{c|}{} & $\pi/18$ \\
        \cline{1-1} \cline{4-4}
        \(\sigma_{s\varphi}\) & \multicolumn{2}{c|}{} & $\pi/9$ \\
        \hline
        \(p_3\) & \multicolumn{3}{c|}{0.5} \\
        \hline
        \(p_4\) & \multicolumn{3}{c|}{0.5} \\
        \hline
        \(p_5\) & \multicolumn{3}{c|}{0.3} \\
        \hline
        \(V_{\textrm{max}}^{r}\) & \multicolumn{3}{c|}{0.05} \\
        \hline
        \(V_{\textrm{max}}^{\theta}\) & \multicolumn{3}{c|}{0.2} \\
        \hline
        \(V_{\textrm{max}}^{\varphi}\) & \multicolumn{3}{c|}{0.4} \\
        \hline
    \end{tabular}
\end{table}

\begin{figure*}
\centering                   
\includegraphics[width=15cm]{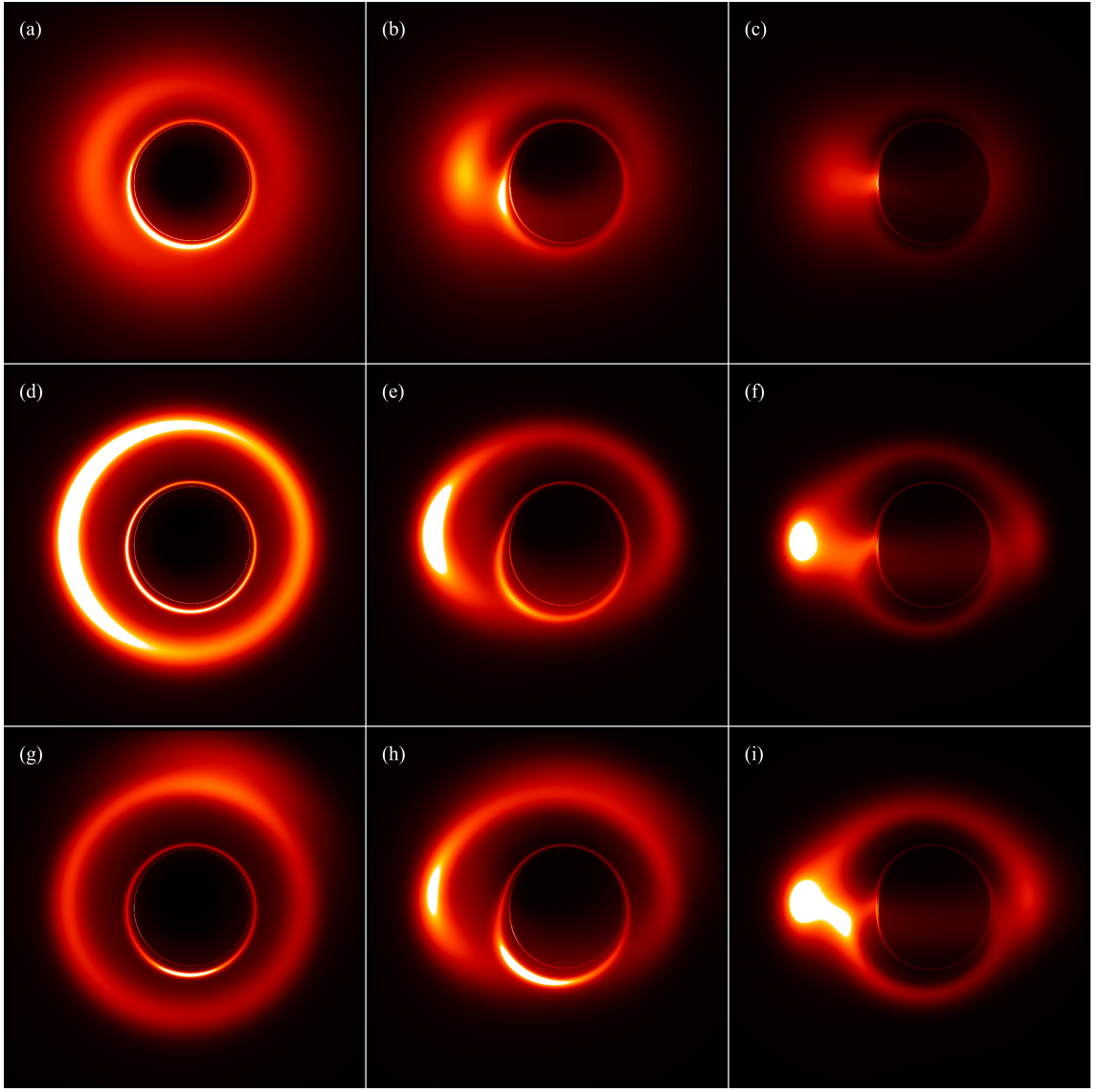}
\caption{Images of a Kerr black hole under different observation perspectives across the environments of Case$_{\textrm{A}}$ (top row), Case$_{\textrm{B}}$ (middle row), and Case$_{\textrm{C}}$ (bottom row). From left to right, the observation inclination angles are $\Theta=17^{\circ}$, $50^{\circ}$, and $80^{\circ}$, respectively, while the observation azimuthal angle is fixed at $\Phi=0^{\circ}$. The parameters for the three accretion environments are listed in table \ref{tab1}. The inner and outer boundaries of the accretion environment are set to $\left(r_{\textrm{in}},r_{\textrm{out}}\right)=\left(r_{\textrm{eh}},500\right)$. In each panel, the lower limit of the color bar is $I_{\textrm{obs}}=0$, and the upper limit is capped at $80\%$ of the maximum $I_{\textrm{obs}}$ in the image. Here, the adaptive step-size parameters are fixed to $h_{\textrm{min}}=0.0025$, $N=1.8$, and $\xi_{\theta}=0.01$, with an RKF56 tolerance of $10^{-8}$.} 
\label{fig4}                 
\end{figure*}

The parameters for the three scenarios are listed in table \ref{tab1}, and the corresponding images are displayed in figure \ref{fig4}. In panel (a), the black hole shadow manifests as an inner shadow, bounded by two highly asymmetric and prominent higher-order photon rings, with the inner one corresponding to the critical curve. Adjacent to these higher-order rings, a moderately bright halo ring is discernible, originating from the radiation compensation. As the observation inclination increases, the Doppler beaming effect induced by azimuthal motion becomes pronounced, yielding two bright spots on the left side of the image. Concurrently, the outermost halo ring deforms into a quasi-elliptical shape. At an inclination of $\Theta=80^{\circ}$, the interior of the critical curve exhibits almost zero intensity, featuring only faint band-like structures. Nevertheless, a bright spot remains discernible on the left side of the critical curve. Building upon Case$_{\textrm{A}}$, the inclusion of the Gaussian bump $j_{\textrm{b}}\left(r,\theta\right)$ introduces an additional bright ring to the image, where the bright spots along this ring exhibit varying morphologies depending on the observation inclination. Furthermore, upon introducing a localized hotspot at $\left(r_\textrm{s},\theta_\textrm{s},\varphi_\textrm{s}\right)=\left(11,90^{\circ},160^{\circ}\right)$, extra bright regions emerge in the image, encompassing both the direct emission and the strongly lensed emission of the hotspot. Overall, \texttt{SpinOctopus} accurately captures key relativistic features such as the critical curve and the black hole shadow. Moreover, owing to the high degrees of freedom within the accretion environment, the images exhibit multiple bright arcs and spots across certain parameter spaces, a characteristic rarely observed in standalone accretion disk scenarios. Consequently, \texttt{SpinOctopus} provides users with highly versatile imaging scenarios, offering potential guidance for inferring high-energy astrophysical environments from observed image features.

\begin{figure*}
\centering                   
\includegraphics[width=15cm]{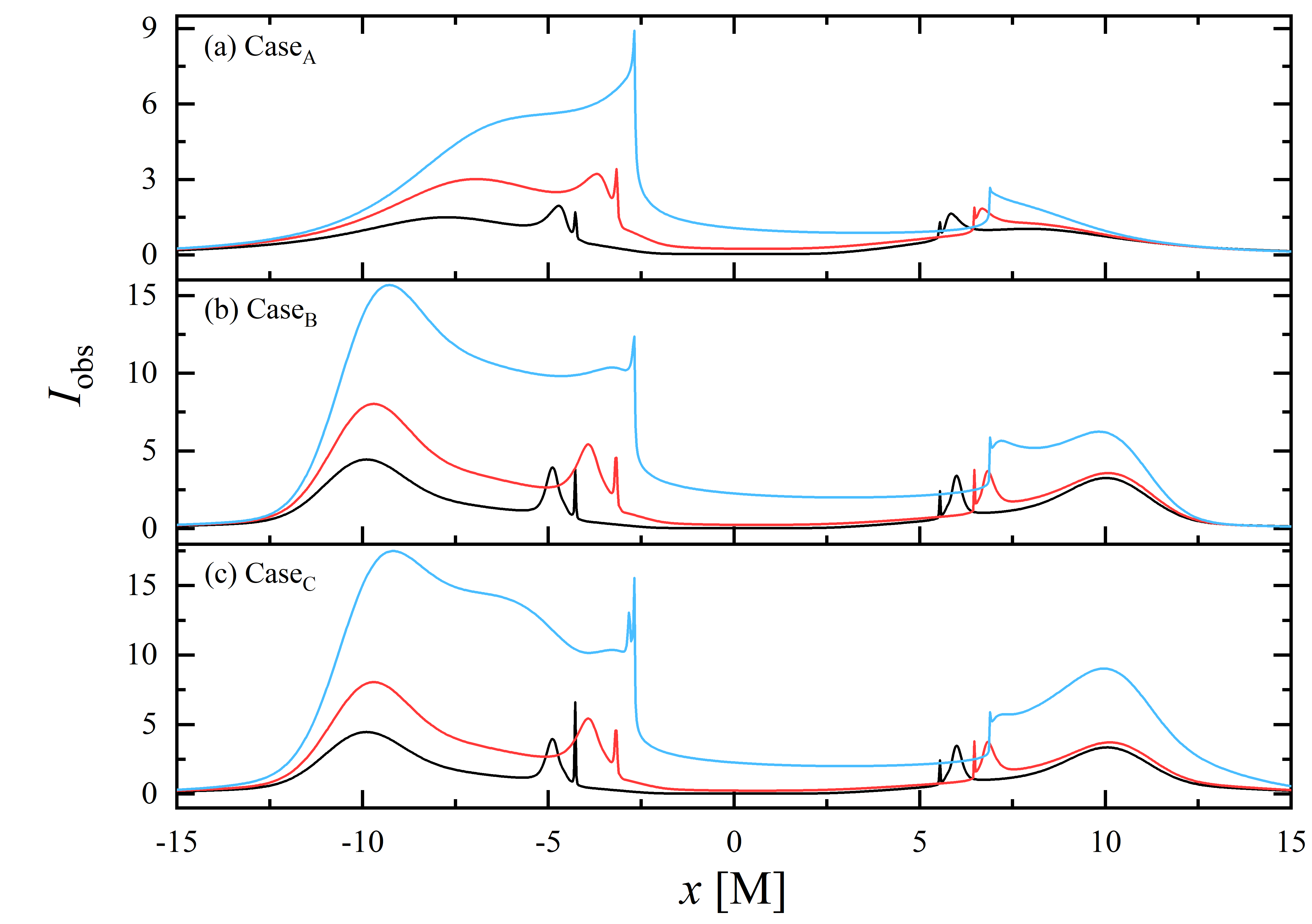}
\caption{One-dimensional specific intensity profiles on the observation screen for different scenarios. In each panel, the black, red, and blue curves correspond to observation inclination angles of $\Theta=17^{\circ}$, $50^{\circ}$, and $80^{\circ}$, respectively.} 
\label{fig5}                 
\end{figure*}

\texttt{SpinOctopus} is also capable of simulating the one-dimensional (1D) specific intensity profiles across the observation screen. Figure \ref{fig5} presents the 1D specific intensity profiles corresponding to Case$_{\textrm{A}}$, Case$_{\textrm{B}}$, and Case$_{\textrm{C}}$ at various observation angles, holding $y=0$ fixed. These profiles offer a clearer visualization of the respective contributions from different components within the accretion environment; for instance, both the radiation compensation and the Gaussian bump elevate the intensity curves. However, a comparison between panel (b) and panel (c) reveals that the localized compact hotspot exerts a marginal impact on the 1D intensity profile. As the observation inclination angle increases, the flux depression region noticeably shifts to the right, corresponding to the deformation of the black hole shadow induced by the frame-dragging effect. Simultaneously, the specific intensity within the shadow region also increases alongside $\Theta$. This behavior arises because the geometric thickness of the accretion environment grows with the radial coordinate, implying a relatively sparse matter distribution in the immediate vicinity of the black hole. Consequently, as the line of sight gradually approaches the equatorial plane, the optical path length of rays traversing the accretion environment increases significantly, thereby contributing to a higher integrated specific intensity.
\subsection{Thin accretion disk images}
In \cite{2021ApJ...918....6C}, the authors derived an analytical model for an optically and geometrically thin accretion disk by fitting time-averaged GRMHD images. \texttt{SpinOctopus} adopts this model to simulate images of geometrically thin accretion disks. In logarithmic space, the emission profile of the accretion disk is formulated as
\begin{equation}\label{67}
\ln\left[\mathcal{F}_{\textrm{D}}\left(r\right)\right] = p_{1}\ln\left(\frac{r}{r_{\textrm{in}}}\right) + p_{2}\left[\ln\left(\frac{r}{r_{\textrm{in}}}\right)\right]^{2}.
\end{equation}
Here, analogous to equation \eqref{47}, $p_{1}$ and $p_{2}$ govern the radial decay of the accretion disk. Specifically, they are dependent on the observation frequency: $p_{1}=-2$ and $p_{2}=-0.5$ correspond to 230 GHz, while $p_{1}=0$ and $p_{2}=-0.75$ correspond to 86 GHz \cite{2021ApJ...918....6C}.

From a numerical perspective, a geometrically thin accretion disk possesses zero thickness; thus, it is unnecessary to integrate the radiative transfer equation to derive the specific intensity of the ray. Instead, we monitor the ray trajectory. Once the coordinates of the ray at two adjacent integration steps simultaneously satisfy
\begin{equation}\label{68}
r_{\textrm{in}} < r_{n+1} < r_{\textrm{out}}
\end{equation}
and
\begin{equation}\label{69}
\cos\theta_{n}\cos\theta_{n+1} \leq 0,
\end{equation}
it indicates that the ray has intersected the accretion disk located at the equatorial plane. Subsequently, the midpoint $\left(r_{n}+r_{n+1}\right)/2$ is approximated as the intersection point $r_{\textrm{hit}}$, which is then substituted into equation \eqref{67} to evaluate the specific emission intensity $\mathcal{F}_{\textrm{D}}\left(r_{\textrm{hit}}\right)$ at the disk. Consequently, the total specific intensity carried by the entire ray is the sum of the intensities acquired during each disk-crossing event:
\begin{equation}\label{70}
I_{\textrm{obs}} = \sum_{i=1}^{\mathcal{N}}f_{n}\mathcal{F}_{\textrm{D}}\left(r^{i}_{\textrm{hit}}\right)g^{i}.
\end{equation}
Here, $\mathcal{N}$ denotes the total number of times the ray intersects the disk. In our code, this number is capped by an upper limit $\mathcal{N}_{\textrm{max}}$, with $\mathcal{N}_{\textrm{max}} \leq 4$. The ray integration is immediately terminated once $\mathcal{N}$ reaches $\mathcal{N}_{\textrm{max}}$. The variables $r^{i}_{\textrm{hit}}$ and $g^{i}$ represent the radial coordinate of the intersection point and the redshift factor of the ray for the $i$-th disk-crossing event, respectively. To fine-tune the image details, an artificial parameter $f_{n}$ is introduced, satisfying $f_{n}=1$ for $i=1$, and $f_{n}=2/3$ for $1 < i \leq \mathcal{N}$ \cite{2021ApJ...918....6C}. Furthermore, to calculate the redshift factor, we adopt the dynamical model of the geometrically thick accretion disk introduced in the previous section, while strictly maintaining $\theta=\pi/2$; that is, $\hat{v}^{\theta}\left(r\right)$ in equation \eqref{53} is constantly set to zero.

\begin{figure*}
\centering                   
\includegraphics[width=15cm]{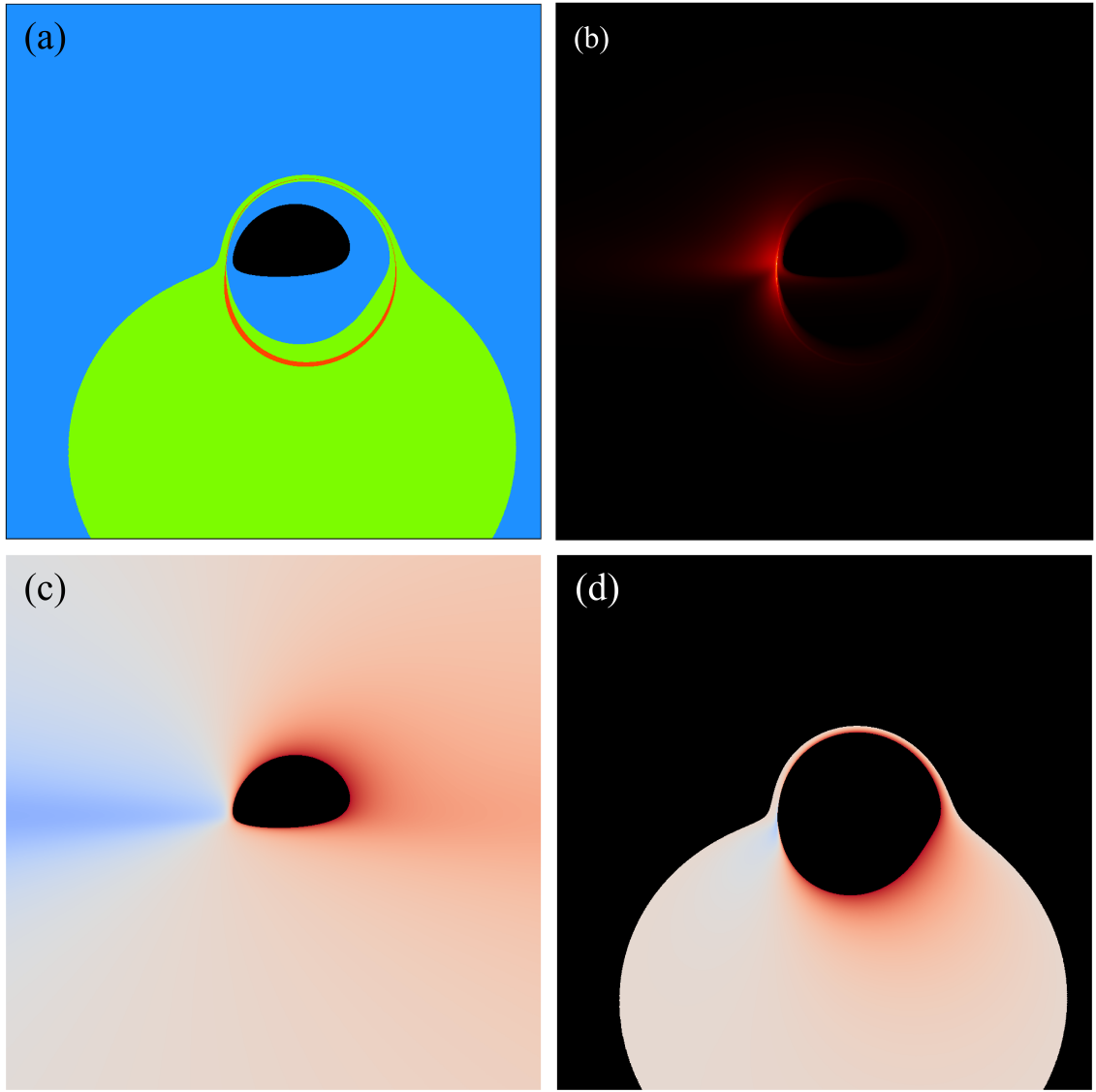}
\caption{Distributions of the disk-crossing number $\mathcal{N}$ [panel (a)], the observed specific intensity [panel (b)], the first-order redshift factor $g^{1}$ [panel (c)], and the second-order redshift factor $g^{2}$ [panel (d)] across the observation screen for a geometrically thin accretion disk. The observation inclination and azimuthal angles are fixed at $\Theta=80^{\circ}$ and $\Phi=0^{\circ}$, respectively. In panel (a), the black, blue, green, and red regions indicate that the rays corresponding to those pixels cross the accretion disk 0, 1, 2, and 3 times, respectively. In panel (b), a gamma color scale with an index of $0.5$ is applied, and the upper limit of the color bar is capped at $80\%$ of the maximum intensity within the screen. In panels (c) and (d), the redshift factor $g$ is visualized using a continuous linear colormap, where red indicates redshift ($g < 1$) and blue indicates blueshift ($g > 1$). For the numerical simulation, the parameters are fixed as $\mathcal{N}_{\textrm{max}}=3$, $p_{1}=-2$, $p_{2}=-0.5$, $V_{\textrm{max}}^{r}=0.01$, $V_{\textrm{max}}^{\varphi}=p_{3}=0.5$, $p_{5}=0.05$, and $V_{\textrm{max}}^{\theta}=0$. The adaptive step-size strategy is identical to that used in figure \ref{fig4}.} 
\label{fig6}                 
\end{figure*}

Adopting an screen grid identical to that of figure \ref{fig4}, we simulate the distributions of the crossing number $\mathcal{N}$, the first-order redshift factor $g^{1}$, the second-order redshift factor $g^{2}$, and the resulting image at an inclination of $\Theta=80^{\circ}$, as displayed in figure \ref{fig6}. The distribution of $\mathcal{N}$ aligns perfectly with theoretical expectations: as $\mathcal{N}$ increases, the illuminated region asymptotically approaches the critical curve, whereas $\mathcal{N}=0$ strictly corresponds to the inner shadow of the black hole. In panel (c), owing to the Doppler effect, the left side of the image is blueshifted while the right side is redshifted. However, this pronounced left-right asymmetry becomes significantly diminished in the second-order redshift factor, as illustrated in panel (d). Furthermore, a prominent redshift is observed in the vicinity of the inner shadow; this is attributed to the substantial radial velocity of the accreting matter coupled with the intense gravitational redshift near the black hole. The underlying distribution pattern of the redshift factors directly dictates the observed image features. For instance, as shown in panel (b), it drives the prominent brightness asymmetry between the left and right sides of the image. Simultaneously, because the geometric thickness of the accretion disk is negligible, the bright band bisecting the shadow (appearing as a faint horizontal line near $y=0$) is exceptionally thin when compared to the images of geometrically thick accretion disks.
\subsection{Tilted thin accretion disk images}
Theoretically, when the angular momentum of the accreting matter is misaligned with the spin axis of the black hole, a tilted accretion disk can form . Intriguingly, based on 17 years of observational data, \cite{2023Natur.621..711C} discovered a periodic precession in the jet of M87$^{*}$, strongly implying that its accretion disk is tilted. \texttt{SpinOctopus} is equipped to simulate the shadows of black holes under the assumption of such a tilted accretion disk. It is important to clarify that, in this context, we do not treat the accretion disk as a light source with an intrinsic emission function. Instead, we adopt a purely geometric assumption, positing a persistent, non-emitting plane at a specific inclination angle. This configuration provides a highly efficient framework for investigating the obscuration effects of a tilted accretion disk on the black hole shadow \cite{Hu:2024imp,Hu:2026rqa}.
\begin{figure*}
\centering                   
\includegraphics[width=8cm]{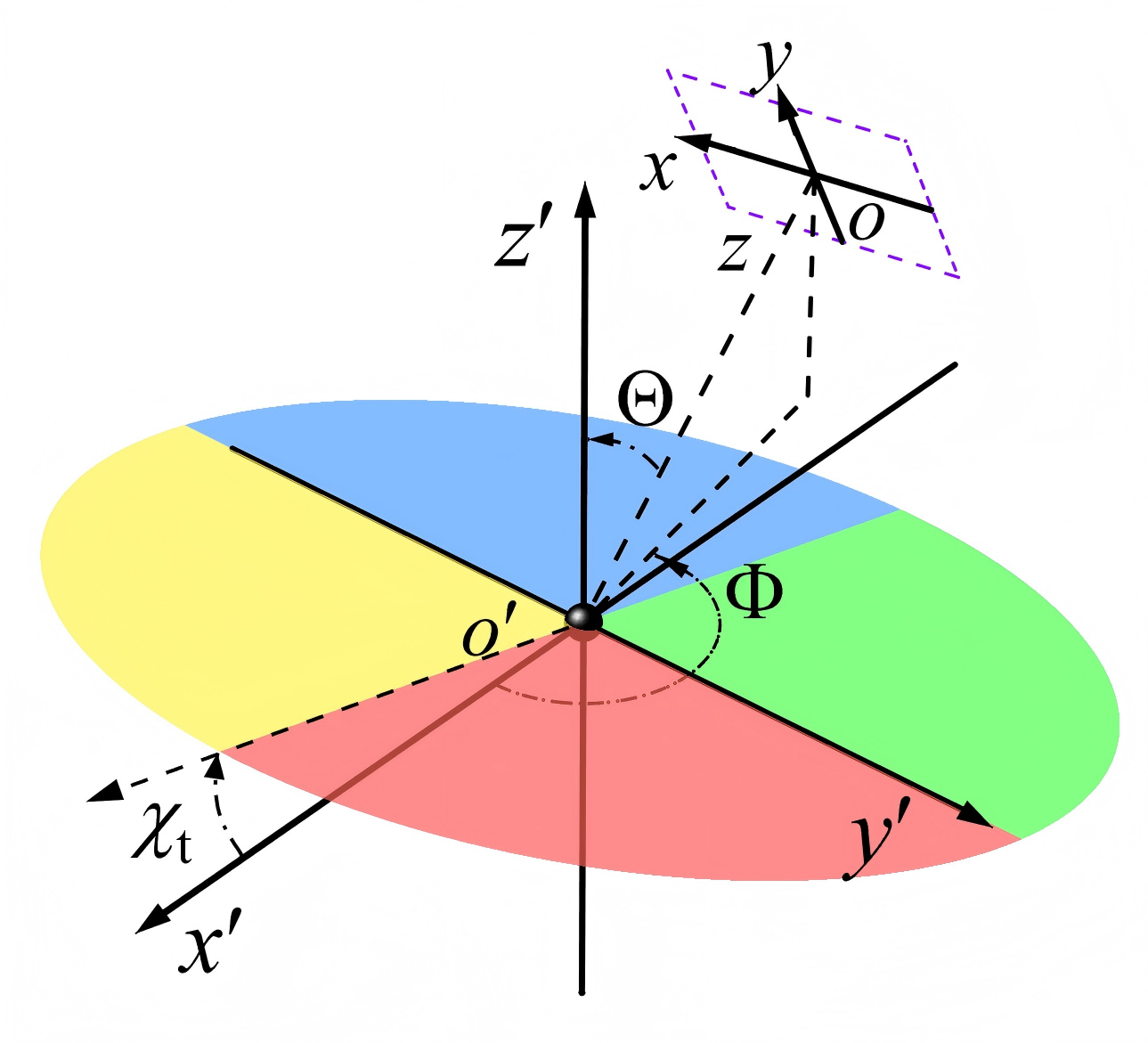}
\caption{Schematic diagram illustrating a geometrically thin tilted accretion disk alongside the observer's local coordinate system and the curved spacetime local coordinate system. The tilt angle of the accretion disk is denoted as $\chi_{\textrm{t}}$, defined such that the region with $x^{\prime}>0$ corresponds to $z^{\prime}>0$. The accretion disk is divided into four color-coded quadrants to track emission from different regions. Other details in the figure are identical to those in figure \ref{fig1}.}      
\label{tilteddisk}              
\end{figure*}

As illustrated in figure \ref{tilteddisk}, the curved spacetime local coordinate system and the observer's local coordinate system remain identical to those defined in figure \ref{fig1}. Here, the tilted accretion disk is generated by rotating the equatorial plane $x^{\prime}o^{\prime}y^{\prime}$ around the $y^{\prime}$-axis by an angle $\chi_{\textrm{t}}$, referred to as the disk tilt angle. The disk is segmented into four distinct regions, colored red, green, blue, and yellow, corresponding to the quadrants defined by $x^{\prime} > 0 \cap y^{\prime} > 0$, $x^{\prime} < 0 \cap y^{\prime} > 0$, $x^{\prime} < 0 \cap y^{\prime} < 0$, and $x^{\prime} > 0 \cap y^{\prime} < 0$, respectively.

In the presence of a tilted accretion disk, rays emitted from the observation screen experience one of three distinct propagation fates, which ultimately dictate the outcomes of the ray-tracing. When a ray directly plunges into the central compact object without intersecting the accretion disk, the corresponding pixel is rendered in black, representing the black hole shadow. Alternatively, if the coordinates of the ray at two successive integration steps satisfy
\begin{equation}\label{td:hitcondition}
\left(x^{\prime}_{n+1}\tan\chi_{\textrm{t}}-z^{\prime}_{n+1}\right)\left(x^{\prime}_{n}\tan\chi_{\textrm{t}}-z^{\prime}_{n}\right) \leq 0,
\end{equation}
it indicates that the ray has struck the tilted accretion disk. Here, $n+1$ and $n$ denote the integration steps, while the coordinates $x^{\prime}$ and $z^{\prime}$ are obtained via equations \eqref{81}--\eqref{83}. Upon impact, the pixel is populated with the color corresponding to the quadrant of the intersection point, and the ray integration is immediately halted. Finally, if the ray successfully escapes to infinity without intersecting either the event horizon or the accretion disk, the associated pixel is colored orange. It is worth noting that the introduction of a disk tilt angle breaks the inherent axisymmetric nature of the spacetime configuration, enabling observers at different azimuthal angles to perceive entirely distinct imaging features.

\begin{figure*}
\centering                   
\includegraphics[width=14cm]{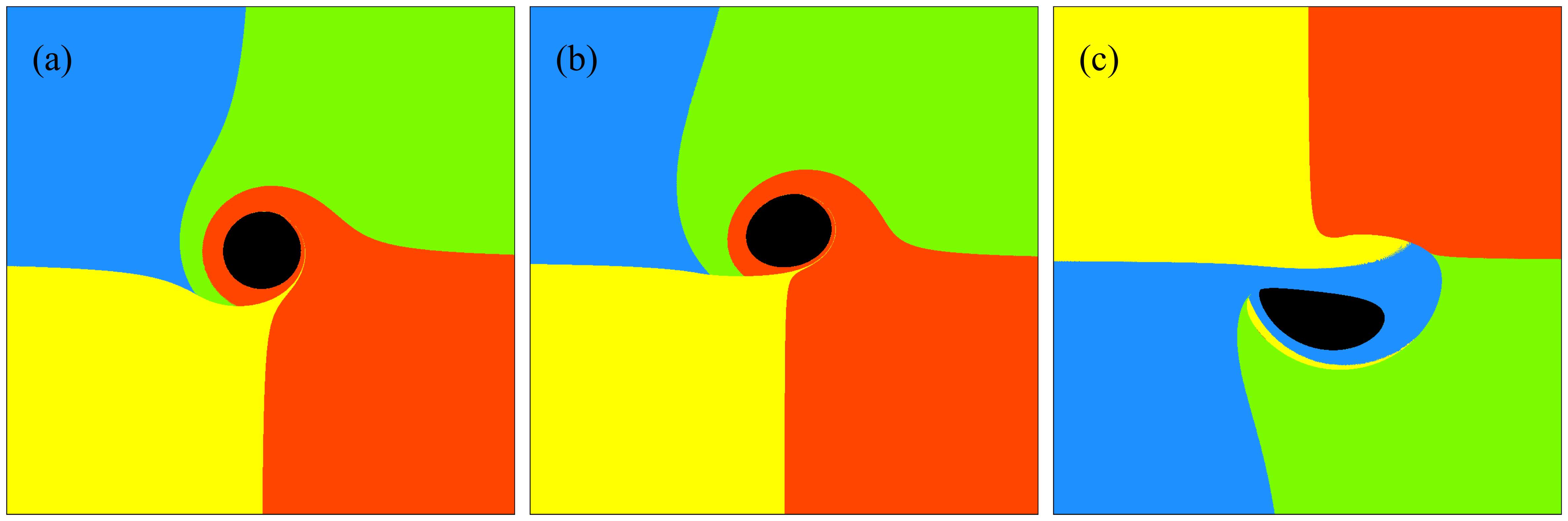}
\caption{Images of a Kerr black hole surrounded by a geometrically thin, tilted accretion disk. From left to right, the observation inclination angles are $\Theta=17^{\circ}$, $50^{\circ}$, and $80^{\circ}$, respectively. The disk tilt angle is fixed at $\chi_{\textrm{t}}=30^{\circ}$, and the observation azimuthal angle is fixed at $\Phi=0^{\circ}$. The adaptive step-size parameters are set to $h_{\textrm{min}}=0.001$, $N=1.8$, and $\xi_{\theta}=0.01$, with an RKF56 tolerance of $10^{-8}$.}  
\label{tdimage}              
\end{figure*}

Setting the field of view to $x,y \in [-10,10]$ with a resolution of $1000 \times 1000$ pixels, we present images of a black hole with a tilted accretion disk ($\chi_{\textrm{t}}=30^{\circ}$) across various observation angles, as shown in figure \ref{tdimage}. It is evident that the tilted accretion disk can not only erode the black hole shadow but also displace its overall position. This alteration is primarily attributed to the elevated disk, which intercepts rays that would have otherwise plunged into the black hole, while simultaneously creating new windows for other rays to fall in. For a more comprehensive discussion on the intriguing imaging features associated with tilted accretion disks, we refer the reader to \cite{Hu:2023pyd,Hu:2024imp,Hu:2026rqa,Chatterjee:2020eqc,White:2020cos}.
\subsection{Spherical accretion flow images}
When the accretion rate of a black hole is significantly below the Eddington limit, as observed in low-luminosity active galactic nuclei (LLAGNs), the cooling timescale of the surrounding fluid vastly exceeds the free-fall timescale, thereby preventing the formation of an accretion disk. Consequently, the accreting matter can only radially accrete onto the black hole in a nearly spherically symmetric manner, giving rise to the classical Bondi accretion \cite{1952MNRAS.112..195B,2004NewAR..48..843E}. In this scenario, the spatial distribution of the accreting matter can be approximated as a spherical shell with non-negligible thickness. \texttt{SpinOctopus} is fully equipped to simulate black hole images under this specific configuration.

Defining the inner and outer radii of the spherical shell as $r_{\textrm{in}}$ and $r_{\textrm{out}}$, respectively, we employ a Gaussian distribution to characterize the emission within the shell:
\begin{equation}\label{71}
\mathcal{F}_{\textrm{S}}\left(r\right) = \textrm{e}^{-\frac{\left(r-r_{\textrm{shell}}\right)^{2}}{2\sigma_{\textrm{shell}}^{2}}},
\end{equation}
where $r_{\textrm{shell}}$ denotes the central radius of the shell, and $\sigma_{\textrm{shell}}$ governs the radial width of the emission. By introducing a weight factor $j_{0}$, the total emission is given by
\begin{equation}\label{72}
j_{\nu}\left(r\right)=j_{0}\mathcal{F}_{\textrm{S}}\left(r\right).
\end{equation}

Regarding the dynamics, although the underlying premise is that matter radially accelerates toward the black hole, we introduce a braking effect into the velocity profile to suppress the acceleration within the spherical shell. This qualitatively mimics the kinematic suppression experienced by matter when traversing a high-density region. Consequently, the local radial velocity $\hat{v}^{r}$ is formulated as
\begin{equation}\label{73}
\hat{v}^{r}\left(r\right) = -V_{\textrm{max}}^{\textrm{shell}}\left(\frac{r_{\textrm{eh}}}{r}\right)^{p_{6}}\textrm{e}^{1-\delta_{\textrm{brake}}\mathcal{F}_{\textrm{S}}\left(r\right)}.
\end{equation}
Here, $V_{\textrm{max}}^{\textrm{shell}}$ designates the maximum radial velocity of the matter; $p_{6}$ governs the rate of velocity variation and is typically set to 0.5; and $\delta_{\textrm{brake}}$ serves as the braking coefficient. A larger $\delta_{\textrm{brake}}$ yields a more pronounced suppression of the velocity within the shell. Adopting the four-velocity framework established in equations \eqref{52}--\eqref{66} and fixing $\hat{v}^{\theta}=\hat{v}^{\varphi}=0$, we obtain the four-velocity $\left(\dot{x}^{t},\dot{x}^{r},0,0\right)$ of the accreting matter derived via the ZAMO projection.

\begin{figure*}
\centering                   
\includegraphics[width=15cm]{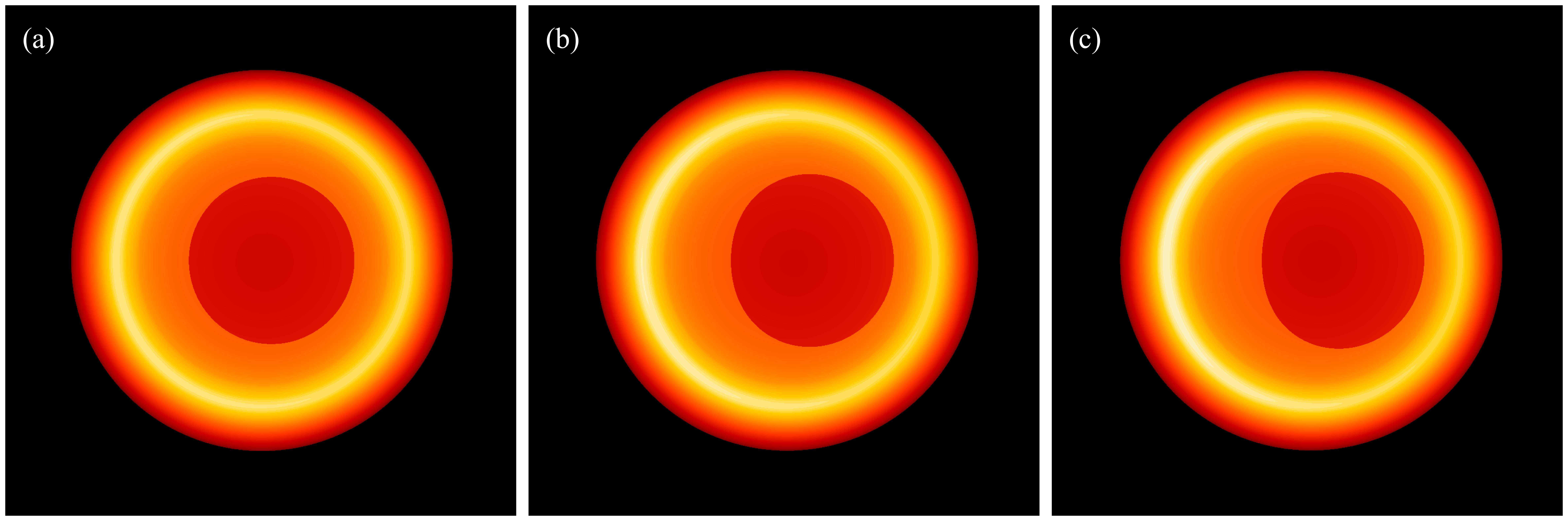}
\caption{Images of a Kerr black hole under a spherical shell accretion environment at various observation inclination angles. From left to right, $\Theta=17^{\circ}$, $50^{\circ}$, and $80^{\circ}$, respectively. In each panel, a gamma color scale with an index of $0.25$ is adopted, and the upper limit of the color bar is capped at $9$. The adaptive step-size strategy for ray integration remains identical to that in figure \ref{fig4}.} 
\label{fig7}                 
\end{figure*}

We position the spherical shell at $\left(r_{\textrm{in}},r_{\textrm{shell}},r_{\textrm{out}}\right)=\left(6,8,10\right)$ and fix the parameters as $V_{\textrm{max}}^{\textrm{shell}}=0.2$, $p_{6}=0.5$, $\delta_{\textrm{brake}}=0.5$, and $\sigma_{\textrm{shell}}=1$. The resulting Kerr black hole images at different observation inclination angles are presented in figure \ref{fig7}. We observe that a distinct bright ring emerges in each panel, contributed by the emission peak within the spherical shell. Notably, the brightness on the left side of this ring intensifies as the observation inclination angle increases. Furthermore, it is found that the outline of the black hole shadow perfectly coincides with the critical curve. This is physically expected, since in a spherically symmetric accretion environment, the observable shadow of a black hole serves as a robust spacetime fingerprint, depending solely on the observation inclination and the intrinsic spacetime parameters \cite{2019ApJ...885L..33N}. A more detailed analysis regarding the influence of shell emission and dynamical parameters on the imaging features will be presented in future work.
\subsection{Jet images}
Astrophysical black holes can launch highly collimated, relativistic plasma jets along their spin axes via the Blandford-Znajek or Blandford-Payne mechanisms \cite{1977MNRAS.179..433B,1982MNRAS.199..883B}. Such structures can serve as prominent light sources illuminating the central black hole \cite{2022A&A...665A.139S,2021ApJ...909..168K,Papoutsis:2022kzp,Zhang:2024lsf}. In \texttt{SpinOctopus}, the jet is modeled as a solid conical structure coaxial with the black hole's spin axis, enabling the qualitative simulation of jet images based on this geometric setup.

We postulate that the jet emission originates from a base radius $r_{\textrm{base}}$ and decays radially outward along the poles. In the angular direction, the jet possesses an initial opening angle at the base, undergoes an expansion phase, and eventually approaches a constant asymptotic opening angle. An emission function that satisfies these geometric properties is formulated as
\begin{equation}\label{74}
j_{\nu}\left(r,\theta\right) = j_{0}j_{\textrm{r}}\left(r\right)j_{\theta}\left(r,\theta_{\textrm{pole}}\right),
\end{equation}
where $j_{0}$ scales the overall emission intensity, while $j_{\textrm{r}}\left(r\right)$ and $j_{\theta}\left(r,\theta_{\textrm{pole}}\right)$ represent the radial and angular distributions, respectively, which are given by
\begin{eqnarray}
j_{\textrm{r}}\left(r\right) &=& \left(\frac{r_{\textrm{base}}}{r}\right)^{p_{7}}, \label{75} \\
j_{\theta}\left(r,\theta_{\textrm{pole}}\right) &=& \textrm{e}^{-\frac{\theta_{\textrm{pole}}^{2}}{2\sigma_{j}^{2}\left(r\right)}}. \label{76}
\end{eqnarray}
Here, $\theta_{\textrm{pole}}=\textrm{Min}[\theta,\pi-\theta]$ ensures the presence of bipolar jets, and $\sigma_{j}\left(r\right)$ takes the form
\begin{equation}\label{77}
\sigma_{j}\left(r\right) = \vartheta_{\infty}-\left(\vartheta_{\infty}-\vartheta_{\textrm{base}}\right)\textrm{e}^{-\frac{r}{\delta_{j}}}.
\end{equation}
The parameter $p_{7}$ dictates the radial decay profile of the jet emission. It is worth noting that in \cite{Papoutsis:2022kzp}, the authors argued that $j_{\textrm{r}}\left(r\right) \propto r^{-5}$ corresponds to millimeter-wavelength emission. However, we retain this index as a free parameter to accommodate a broader range of astrophysical scenarios. The parameters $\vartheta_{\textrm{base}}$ and $\vartheta_{\infty}$ denote the initial opening angle at the jet base and the stable asymptotic opening angle post-expansion, respectively. $\delta_{j}$ characterizes the transition length scale required for the opening angle to evolve from $\vartheta_{\textrm{base}}$ to $\vartheta_{\infty}$, essentially governing the expansion rate of the jet. It must be emphasized that $j_{\textrm{r}}\left(r\right)$ is physically valid only for $r \geq r_{\textrm{base}}$, as the emission vanishes in the region $r < r_{\textrm{base}}$. However, for a jet where $r_{\textrm{eh}} < r_{\textrm{base}}$, the ray-tracing trajectories can still probe the region $r_{\textrm{eh}} < r < r_{\textrm{base}}$, which could introduce numerical anomalies if equation \eqref{75} were blindly applied. Consequently, we explicitly enforce $j_{\textrm{r}}\left(r\right)=0$ for $r < r_{\textrm{base}}$ within the code.

Within our solid jet model, the ejected matter attains its maximum velocity strictly along the spin axis, while the radial velocity of the off-axis matter decays with an increasing polar angle $\theta$. Concurrently, as the jet opening angle saturates, the radial velocity of the jet asymptotically approaches a stable terminal value. A physical radial velocity profile satisfying these kinematic constraints is formulated as
\begin{equation}\label{78}
\hat{v}^{r}\left(r,\theta\right) = \left[V_{\textrm{max}}^{\textrm{jet}}-\left(V_{\textrm{max}}^{\textrm{jet}}-V_{\textrm{base}}\right)\textrm{e}^{-\frac{r-r_{\textrm{base}}}{\delta_{j}}}\right]\textrm{e}^{-\frac{\theta_{\textrm{pole}}^{2}}{2\sigma_{j}^{2}\left(r\right)}},
\end{equation}
where $V_{\textrm{base}}$ denotes the initial ejection velocity at the jet base, and $V_{\textrm{max}}^{\textrm{jet}}$ represents the terminal velocity achievable by the jet. 

In addition to the radial velocity, we introduce an azimuthal velocity component in the plane perpendicular to the spin axis, imparting a gyrating motion with a compact radius to the plasma. This toroidal velocity is expressed as
\begin{equation}\label{79}
\hat{v}^{\varphi}\left(r,\theta\right) = \textrm{sign}\left(\cos\theta\right)r_{\textrm{gyra}}\left[\frac{\sigma_{j}\left(r\right)}{\vartheta_{\textrm{base}}}\right]\omega_{\textrm{jet}}\left(\frac{r_{\textrm{base}}}{r}\right)^{p_{8}}\textrm{e}^{-\frac{\theta_{\textrm{pole}}^{2}}{2\sigma_{j}^{2}\left(r\right)}},
\end{equation}
where $r_{\textrm{gyra}}$ specifies the minimum gyration radius of the matter, and $\omega_{\textrm{jet}}$ denotes the angular velocity at the base (which also corresponds to the maximum angular velocity). The parameter $p_{8}$ further dictates the radial decay of this angular velocity. The sign of the azimuthal velocity $\hat{v}^{\varphi}\left(r,\theta\right)$ is determined by the $\textrm{sign}(\cos\theta)$ term, naturally reflecting the anti-symmetric toroidal motion of the matter within the two opposing bipolar jets. Note that setting $\omega_{\textrm{jet}}=0$ seamlessly reduces the model to a purely radially accelerating jet. By substituting $\hat{v}^{r}$ and $\hat{v}^{\varphi}$ into equations \eqref{63}--\eqref{66}, we obtain the four-velocity $\left(\dot{x}^{t},\dot{x}^{r},0,\dot{x}^{\varphi}\right)$ of the matter within the jet region.

\begin{figure*}
\centering                   
\includegraphics[width=15cm]{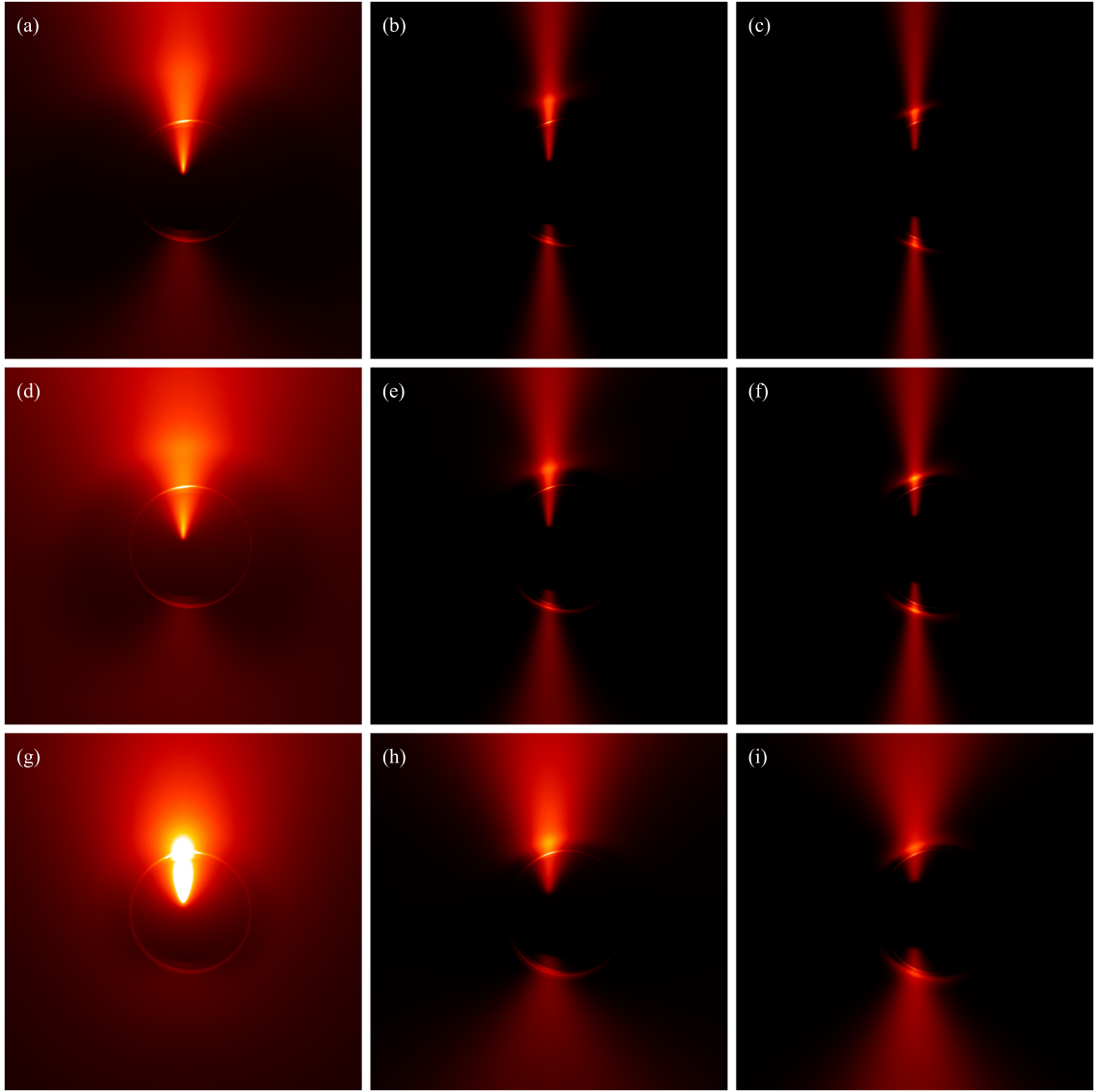}
\caption{Jet images for various observation inclination angles and jet opening-angle parameters. From left to right, the observation angles are $\Theta=17^{\circ}$, $50^{\circ}$, and $80^{\circ}$, respectively. The jet opening-angle parameters $\left(\vartheta_{\textrm{base}},\vartheta_{\infty}\right)$ are set to $\left(0.05,0.15\right)$ for the top row and $\left(0.05,0.26\right)$ for the middle and bottom rows. The transition length scale for the opening-angle expansion is $\delta_{j}=50$ for the top two rows and $\delta_{j}=8$ for the bottom row. Here, the jet base is positioned at $r_{\textrm{base}}=2$, and the radial decay index is $p_{7}=2$. For the kinematic parameters, we fix $V_{\textrm{base}}=0.1$, $V_{\textrm{max}}^{\textrm{jet}}=0.8$, $r_{\textrm{gyra}}=0.01$, $\omega_{\textrm{jet}}=0.15$, and $p_{8}=0.5$. In each panel, a gamma color scale with an index of $0.75$ is applied, and the upper limit of the color bar is capped at $80\%$ of the maximum intensity within the screen. The adaptive step-size strategy remains identical to that in figure \ref{fig4}.}  
\label{fig8}                 
\end{figure*}

Adopting an observation screen grid identical to that in figure \ref{fig4}, we simulate the Kerr black hole images for various observation inclination angles and jet opening-angle evolutions, as illustrated in figure \ref{fig8}. At low observation angles, we find that the complete critical curve emerges only when the jet opening angle reaches a certain threshold, as seen in panels (d) and (g). This phenomenon of critical curve disappearance has also been reported in the dual-cone jet model \cite{Papoutsis:2022kzp}. Simultaneously, a cone-like bright region is discernible at the center of the image. Specifically, when a smaller transition scale $\delta_{j}$ is selected, both the opening angle and the radial velocity of the jet saturate rapidly, rendering this cone-like region exceptionally bright, as demonstrated in panel (g). As the observation inclination angle increases, the projected area of the jet's geometric structure on the observation screen decreases, leading to an overall reduction in the image size. In this high inclination case, the northern and southern jets manifest as two distinct cone shapes in the image, with each side featuring two prominent arc structures. The slender, brighter inner arc constitutes a segment of the critical curve, whereas the broader outer arc originates from the region where the jet velocity has reached saturation. Furthermore, a larger jet opening angle enhances the overall image brightness and exposes a more substantial portion of the critical curve.
\subsection{Point source images}
We now disregard large-scale emitting structures such as accretion disks and jets, shifting our focus exclusively to the imaging features generated by a single point source in the vicinity of the central compact object. This approach offers an elegant avenue to isolate and analyze the gravitational lensing effects inherent to the spacetime. To this end, \texttt{SpinOctopus} provides two distinct simulation schemes. 

In Scheme A, the point source is assumed to be stationary and is treated as a uniformly emitting sphere. The corresponding image is simulated against the backdrop of a celestial sphere. The outcomes of this scheme typically encompass the primary and higher-order lensed images of the point source, the black hole shadow, and image distortions induced by frame-dragging---commonly visualized as the standard four-color map \cite{Bohn:2014xxa}. 

In Scheme B, we prescribe the four-velocity, emission function, and absorption coefficient of the point source, and subsequently simulate its image from a rigorous radiative transfer perspective. Within this framework, the point source serves as the sole emitter, completely devoid of any secondary light sources.
\subsubsection{Scheme A}
\begin{figure*}
\centering                   
\includegraphics[width=12cm]{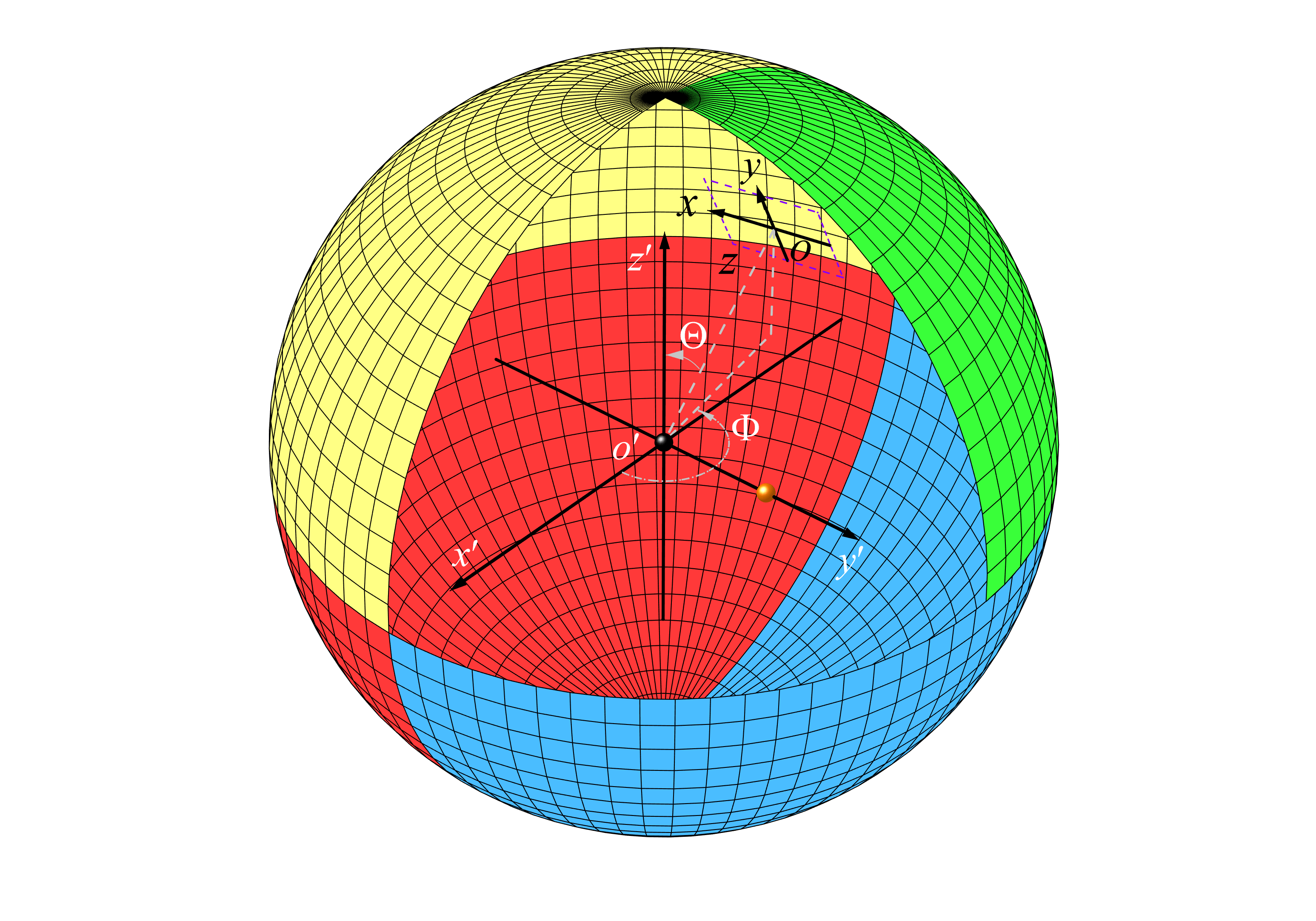}
\caption{Schematic diagram illustrating the celestial sphere background, the point source, and the coordinate systems. Here, the coordinate systems and observation parameters remain strictly identical to those in figure \ref{fig1}. The small orange sphere represents the sample point source.}      
\label{fig9}                 
\end{figure*}
The celestial sphere background and coordinate systems adopted in Scheme A are illustrated in figure \ref{fig9}. The curved spacetime local coordinate system $x^{\prime}y^{\prime}z^{\prime}$, the observer's local coordinate system $xyz$, and the observation parameters remain strictly identical to those defined in figure \ref{fig1}. The primary distinction lies in the introduction of a massive celestial sphere with a radius of $r_{\textrm{cls}}=1200$, which completely encompasses the central compact object, the observer, and the point source (represented by the orange sphere). This celestial sphere is partitioned into four regions based on the conditions $y^{\prime} > 0 \cap z^{\prime} > 0$, $y^{\prime} > 0 \cap z^{\prime} < 0$, $y^{\prime} < 0 \cap z^{\prime} > 0$, and $y^{\prime} < 0 \cap z^{\prime} < 0$. These quadrants are consequently color-coded green, blue, yellow, and red, respectively, under the assumption that each region emits light of the corresponding color.

Against this celestial sphere background, rays traced backward from the observation screen $xoy$ encounter one of three distinct destinies. First, a ray may escape and intersect the celestial sphere without plunging into the black hole or striking the point source. In this case, the corresponding pixel is populated with the color associated with its intersection coordinate on the celestial sphere. Second, a ray may directly plunge into the black hole; the corresponding pixel is then rendered in black. Third, a ray may strike the point source, resulting in the pixel being colored orange. Within the numerical code, we explicitly specify that the ray integration terminates immediately upon the occurrence of either of the latter two events. On the observation screen, the collection of black pixels maps out the black hole shadow, whereas the collection of orange pixels constitutes the primary or gravitationally lensed image of the point source.

It must be emphasized that the so-called ``point source'' is not a geometric singularity of infinitesimal dimensions; instead, it is modeled as a compact sphere with a non-negligible radius, allowing it to serve as a resolvable target during the ray-tracing procedure. In our setup, the radius of the point source is set to $r_{\textrm{source}}=0.5$ (a parameter that can be freely adjusted within the code). A collision between the ray and the point source is registered whenever the ray coordinates at two successive integration steps satisfy the condition
\begin{eqnarray}\label{80}
\left[\sqrt{\left(x_{n+1}^{\prime}-x_{\textrm{s}}^{\prime}\right)^{2}+\left(y_{n+1}^{\prime}-y_{\textrm{s}}^{\prime}\right)^{2}+\left(z_{n+1}^{\prime}-z_{\textrm{s}}^{\prime}\right)^{2}}-r_{\textrm{source}}\right] \nonumber \\
\times \left[\sqrt{\left(x_{n}^{\prime}-x_{\textrm{s}}^{\prime}\right)^{2}+\left(y_{n}^{\prime}-y_{\textrm{s}}^{\prime}\right)^{2}+\left(z_{n}^{\prime}-z_{\textrm{s}}^{\prime}\right)^{2}}-r_{\textrm{source}}\right] \leq 0.
\end{eqnarray}
Here, $\left(x_{\textrm{s}}^{\prime},y_{\textrm{s}}^{\prime},z_{\textrm{s}}^{\prime}\right)$ specifies the center coordinates of the point source. The subscripts $n+1$ and $n$ denote two consecutive integration steps, with the spatial coordinates being derived via equations \eqref{81}--\eqref{83}.

\begin{figure*}
\centering                   
\includegraphics[width=14cm]{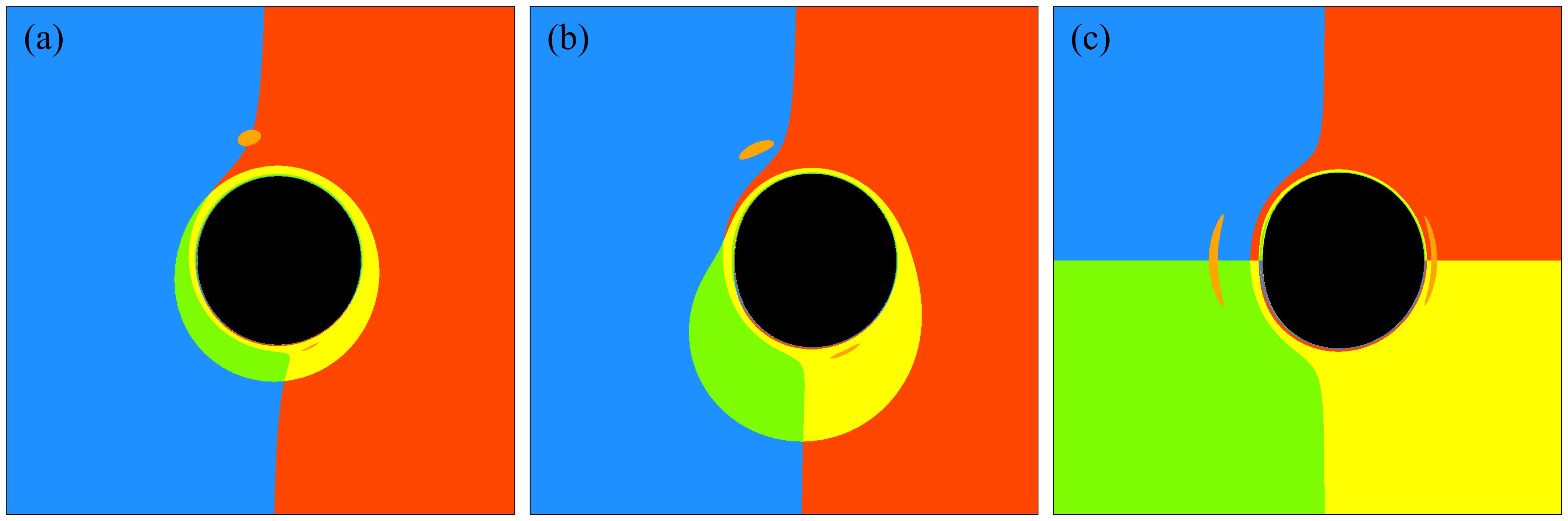}
\caption{Simulated images of a point source at various observation inclination angles. From left to right, the observation angles are $\Theta=30^{\circ}$, $\Theta=60^{\circ}$, and $\Theta=90^{\circ}$, respectively. Here, the observation azimuthal angle is fixed at $\Phi=0^{\circ}$. The integration parameters are set to $h_{\textrm{min}}=0.001$, $N=1.8$, and $\xi_{\theta}=0.01$, with an RKF56 tolerance of $10^{-8}$.}      
\label{fig10}                 
\end{figure*}
Fixing the point source at $\left(x_{\textrm{s}}^{\prime},y_{\textrm{s}}^{\prime},z_{\textrm{s}}^{\prime}\right)=\left(-6,0,0\right)$ and adopting an observation screen grid identical to that in figure \ref{fig4}, we simulate the lensed images across different observation inclination angles, as illustrated in figure \ref{fig10}. At $\Theta=30^{\circ}$, the black hole shadow appears nearly circular, encompassed by a prominent lensing band. The primary image of the point source exhibits an elliptical shape, manifesting slightly to the left near $y=0$ owing to the frame-dragging effect; its secondary image, appearing as a minuscule arc within the yellow segment of the lensing band, is located almost diametrically opposite the primary image with respect to the image center. Meanwhile, the macroscopic distortion of the background celestial sphere visually traces the frame-dragging field. As the observation inclination angle increases to $60^{\circ}$, the left side of the shadow becomes visibly deformed, and its overall contour shifts to the right. Both the primary and secondary images of the point source undergo substantial stretching. When the observer, the black hole, and the point source reach a collinear alignment (i.e., at $\Theta=90^{\circ}$), the lensed images of the point source emerge on opposite sides of the black hole shadow. For a non-spinning or slowly spinning black hole, this exact alignment would produce a complete Einstein ring. However, in our high-spin scenario ($a=0.94$), the intense frame-dragging effect drastically breaks the azimuthal symmetry, disrupting the continuous ring into discrete lensed images.

\begin{figure*}
\centering                   
\includegraphics[width=9cm]{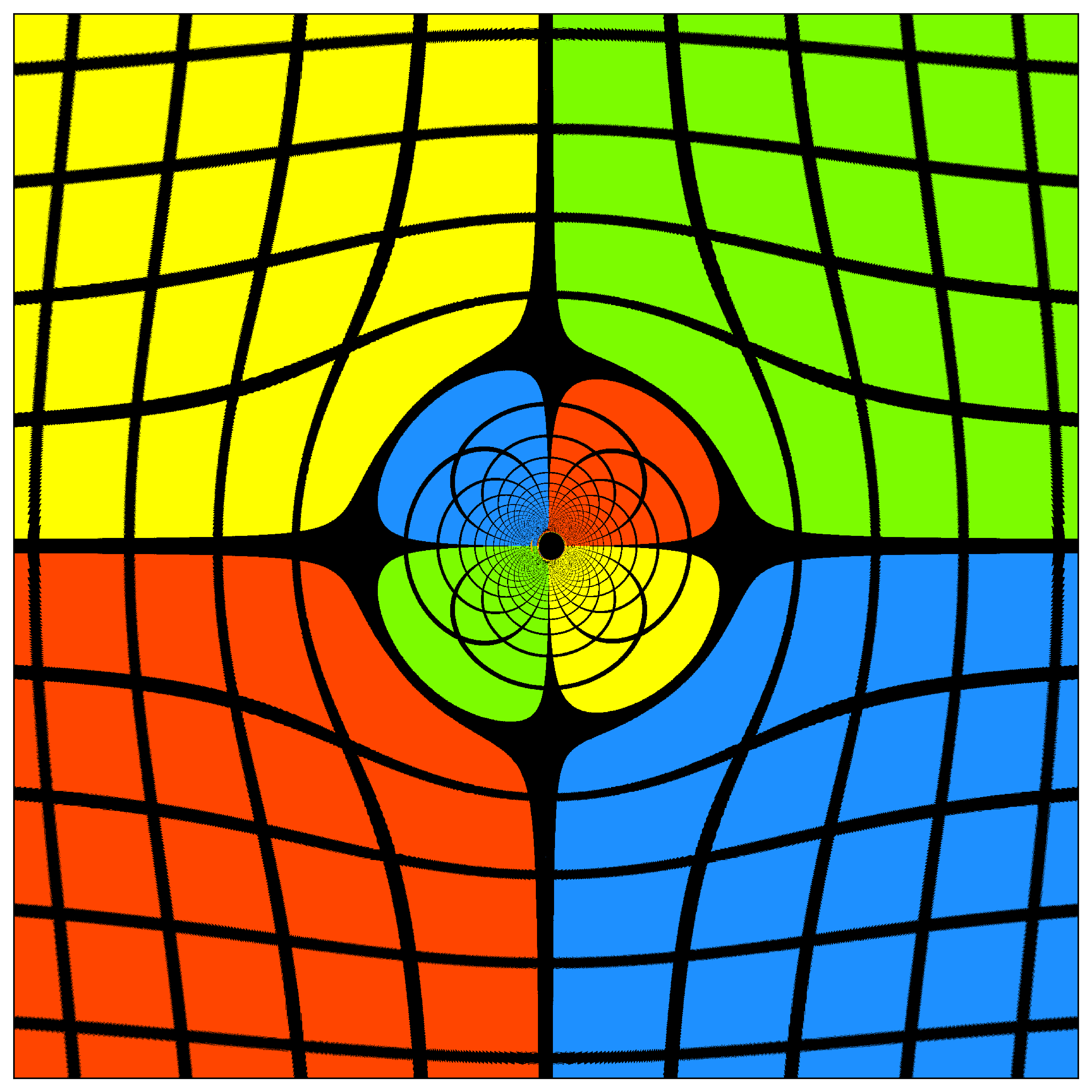}
\caption{Resimulation of the scenario depicted in panel (c) of figure \ref{fig10} under a large field of view. Here, the observation screen is expanded to $x,y \in [-200,200]$ with a resolution of $2000 \times 2000$ pixels, the black coordinate grid corresponds to the longitudinal and latitudinal lines of the celestial sphere.}      
\label{fig11}                 
\end{figure*}
In fact, when simulating the four-color map, \texttt{SpinOctopus} also supports rendering coordinate grids comprising solid black lines of longitude and latitude, as illustrated in figure \ref{fig11}. In this instance, the grid spacing on the celestial sphere is configured to $2^{\circ}$. The distortion of these grid lines provides a more intuitive visualization of the underlying spatial curvature and the gravitational field. However, it is imperative to clarify that, because we assume an observer located at practical infinity, the rays emitted backward from the observation screen are strictly parallel. This geometrical setup is fundamentally distinct from the field-of-view logic underpinning fisheye projection techniques. Consequently, capturing the large-scale spatial distortion effects in this parallel-ray framework necessitates an expansive field of view coupled with a sufficiently high resolution. Within the code, the display of this latitude-longitude grid can be toggled on or off at the user's discretion.
\subsubsection{Scheme B}
We assume that the point source does not emit uniformly; instead, its emission intensity radially decays outward from its core. Accordingly, the emission function of the point source can be modeled using a Gaussian distribution consistent with \cite{Pu:2016eml}, formulated as
\begin{equation}\label{84}
j_{\nu} = j_{0}\textrm{e}^{\left[-\frac{\left(x^{\prime}-x_{\textrm{s}}^{\prime}\right)^{2}+\left(y^{\prime}-y_{\textrm{s}}^{\prime}\right)^{2}+\left(z^{\prime}-z_{\textrm{s}}^{\prime}\right)^{2}}{2\sigma_{\textrm{point}}^{2}}\right]},
\end{equation}
where $\left(x^{\prime},y^{\prime},z^{\prime}\right)$ represents the ray coordinates derived via the transformations \eqref{81}--\eqref{83}, and $\sigma_{\textrm{point}}$ characterizes the internal emission profile of the point source.

Regarding the dynamics of the point source, for the sake of simplicity, we dispense with a continuous temporal or spatial velocity distribution; instead, we assign the velocity for a specific spatial slice. In other words, we directly prescribe an instantaneous physical velocity $\left(\hat{v}^{r},\hat{v}^{\theta},\hat{v}^{\varphi}\right)$ and subsequently compute the four-velocity of the point source via the ZAMO framework to facilitate the subsequent redshift calculations.

\begin{figure*}
\centering                   
\includegraphics[width=14cm]{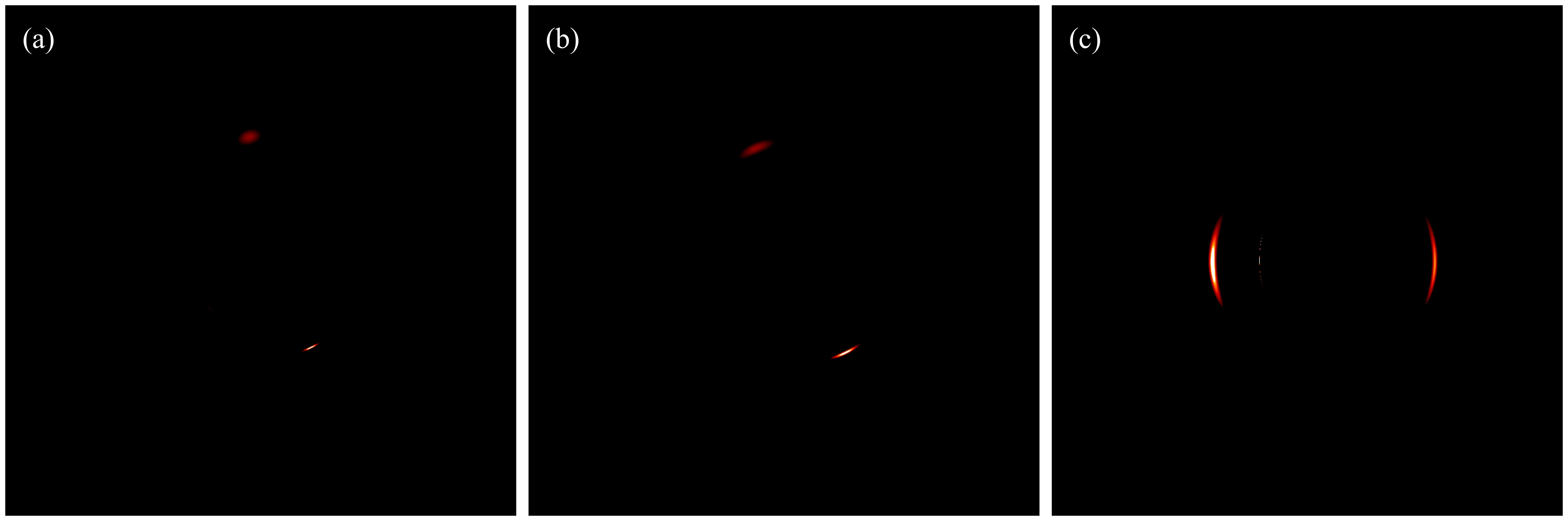}
\caption{Simulated images of the point source at various observation inclination angles using the radiative transfer scheme (Scheme B). From left to right, the observation inclination angles are $\Theta=30^{\circ}$, $60^{\circ}$, and $90^{\circ}$, respectively. Here, the point source, with a radius of $r_{\textrm{source}}=0.5$, is fixed at $\left(x_{\textrm{s}}^{\prime},y_{\textrm{s}}^{\prime},z_{\textrm{s}}^{\prime}\right)=\left(-6,0,0\right)$, and the observation azimuthal angle is fixed at $\Phi=0^{\circ}$. The internal emission profile parameter is set to $\sigma_{\textrm{point}}=0.3$, and the instantaneous physical velocity is prescribed as $\left(\hat{v}^{r},\hat{v}^{\theta},\hat{v}^{\varphi}\right)=\left(0.2,0.05,0.4\right)$. The integration parameters remain identical to those in figure \ref{fig10}. For image visualization, a linear colormap is adopted to represent the observed specific intensity, with the upper limit of the color bar capped at $80\%$ of the maximum intensity within the screen.}      
\label{fig12}                 
\end{figure*}

Proceeding from the geometrical setup in figure \ref{fig10}, we discard the celestial sphere background and simulate the images by incorporating the point source emission governed by equation \eqref{84}; the results are presented in figure \ref{fig12}. We observe that under the illumination of such a compact, localized source, the black hole critical curve is rarely fully manifested, with only a partial arc emerging in panel (c). The resulting visual features consist of both the primary and gravitationally lensed images of the point source. In panels (a) and (b), the lensed images exhibit a higher intensity than the primary image. Conversely, in the case of $\Theta=90^{\circ}$, the image on the left is significantly brighter than the one on the right. This pronounced asymmetry is a direct consequence of Doppler beaming, driven by the substantial azimuthal velocity component of the point source's instantaneous motion.
\subsection{Massive particle orbits and light curves}
Timelike particles in curved spacetime can be conceptualized as dynamic point sources, which continuously emit electromagnetic radiation as they orbit the central compact object. The resulting time-dependent electromagnetic signals captured by distant observers manifest as light curves. Crucially, these light curves encode the intricate orbital dynamics of the massive particles, thereby serving as a powerful probe for deciphering the intrinsic properties of the spacetime and its surrounding high-energy environment \cite{Huang:2024wpj,Tan:2026yjg}. \texttt{SpinOctopus} is fully equipped to integrate timelike particle orbits under arbitrary initial conditions and simulate the corresponding light curves.

To this end, our numerical framework incorporates five distinct initialization schemes for the generalized coordinates and conjugate momenta of the massive particles. In the first scheme, the initial state is prescribed as $\left(0,r^{\textrm{input}},\theta^{\textrm{input}},\varphi^{\textrm{input}},p_{t}^{\textrm{input}},p_{r},0,p_{\varphi}^{\textrm{input}}\right)$, where the superscript ``input'' denotes user-specified parameters, and $p_{r}$ is derived by enforcing the Hamiltonian constraint $\mathcal{H}=-1/2$. The second scheme is analogous to the first, except that we explicitly set $p_{r}=0$ and instead solve for $p_{\theta}$ via $\mathcal{H}=-1/2$. In the third scheme, after specifying the generalized coordinates and the specific angular momentum $p_{\varphi}$, we impose $p_{r}=p_{\theta}=0$ and subsequently determine $p_{t}$ through the Hamiltonian constraint. Note that under the metric signature $\left(-,+,+,+\right)$ adopted in our work, the condition $p_{t} < 0$ inherently holds. 

The fourth scheme entails providing the generalized coordinates alongside the initial physical velocity $\left(\hat{v}^{r},\hat{v}^{\theta},\hat{v}^{\varphi}\right)$. Leveraging the accretion disk dynamical framework, we first calculate the corresponding four-velocity and then evaluate the conjugate momenta via equations \eqref{4}--\eqref{7}. This approach is particularly well-suited for simulating the orbits of accreting particles. It should be clarified, however, that the preceding schemes can also generate plunging accretion orbits, provided the initial conditions are sampled within the inward-sloping plunging region of the effective potential prior to crossing the event horizon. In the fifth scheme, the code bypasses any internal constraint solvers, requiring all eight initial variables to be explicitly supplied by the user. This option caters to scenarios where specific, pre-calculated orbital initial conditions are already available. For instance, one might employ our previously developed rational orbit solver \cite{Li:2026pfa} to extract the initial conditions for rational orbits, which are then fed into \texttt{SpinOctopus} to simulate the corresponding trajectories and light curves.

\begin{figure*}
\centering                   
\includegraphics[width=14cm]{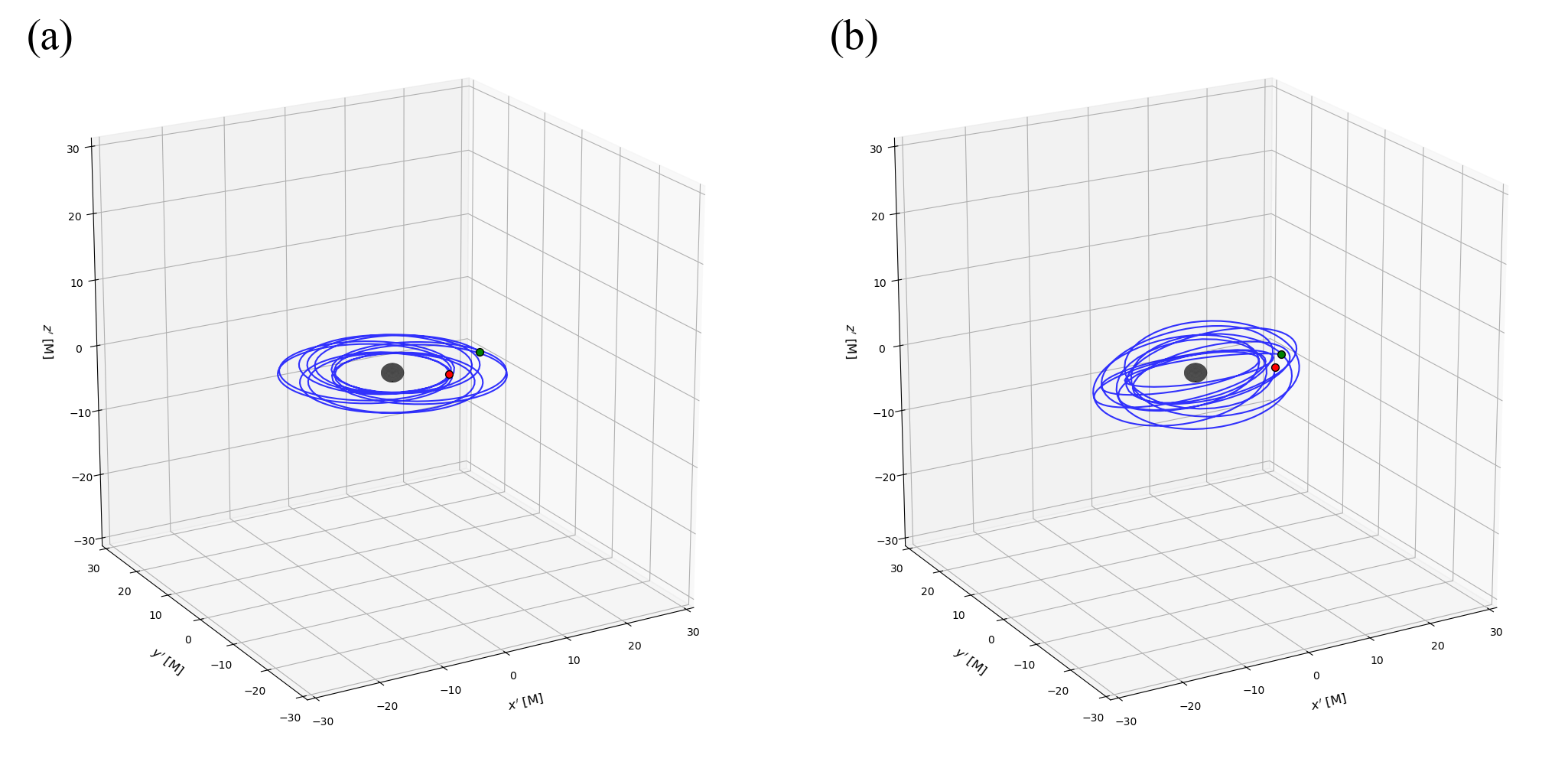}
\caption{Spatial evolution of orbits $\textrm{O}_{1}$ (left panel) and $\textrm{O}_{2}$ (right panel). The green and red spheres indicate the starting and terminating points of the trajectories, respectively. Here, the integration step-size for the RK6 method is set to 0.1, and the total proper time elapsed for the particle motion is 2500.}      
\label{fig13}                 
\end{figure*}
Once the initial conditions of the massive particle are specified, \texttt{SpinOctopus} employs a RK6 integrator to simulate the orbit, yielding the canonical variables of the particle at any arbitrary time. Subsequently, by introducing the parameter $r_{\textrm{source}}$, the continuous trajectory is discretized into a sequence of small spheres; effectively, the orbit is modeled as an ensemble of point sources. During the ray-tracing procedure, any intersection between a ray and a sphere contributes to the photon count (or count rate) recorded at the observation time $T_{\textrm{obs}}=t_{\textrm{p}}+t_{\textrm{m}}$. Here, $t_{\textrm{p}}$ represents the coordinate time of flight required for the ray to strike the sphere, while $t_{\textrm{m}}$ denotes the coordinate time elapsed for the massive particle to reach that specific spherical location. Upon the completion of the ray-tracing, the distribution of the photon count (or count rate) as a function of $T_{\textrm{obs}}$ naturally constitutes the light curve corresponding to the timelike orbit. It is worth noting that, to prevent false positives in detecting collisions between the rays and the spheres, we stipulate that a valid intersection is registered only when the ray exits the sphere. Mathematically, this requires the ray coordinates at two successive integration steps to satisfy
\begin{eqnarray}
\left(x^{\prime}_{n+1}-x^{\prime}_{i}\right)^{2} + \left(y^{\prime}_{n+1}-y^{\prime}_{i}\right)^{2} + \left(z^{\prime}_{n+1}-z^{\prime}_{i}\right)^{2} \geq r_{\textrm{source}}^{2} \label{85}
\end{eqnarray}
and
\begin{eqnarray}
\left(x^{\prime}_{n}-x^{\prime}_{i}\right)^{2} + \left(y^{\prime}_{n}-y^{\prime}_{i}\right)^{2} + \left(z^{\prime}_{n}-z^{\prime}_{i}\right)^{2} \leq r_{\textrm{source}}^{2}, \label{86}
\end{eqnarray}
where $\left(x^{\prime}_{i},y^{\prime}_{i},z^{\prime}_{i}\right)$ denotes the coordinates of the $i$-th sphere along the discretized trajectory.

Fixing the initial generalized coordinates of the timelike particle at $\left(t,r,\theta,\varphi\right)=\left(0,15,\pi/2,0\right)$, and specifying its specific energy and specific angular momentum as $0.96$ and $3.55$, respectively, we utilize the first and second initialization schemes to compute two distinct timelike orbits, denoted as $\textrm{O}_{1}$ and $\textrm{O}_{2}$, as illustrated in figure \ref{fig13}. Orbit $\textrm{O}_{1}$ (left) is a stable equatorial bound orbit. In contrast, orbit $\textrm{O}_{2}$ (right) exhibits pronounced precession in the $\theta$-direction, arising from the non-zero initial momentum $p_{\theta}$ specified in its initial conditions.

\begin{figure*}
\centering                   
\includegraphics[width=15cm]{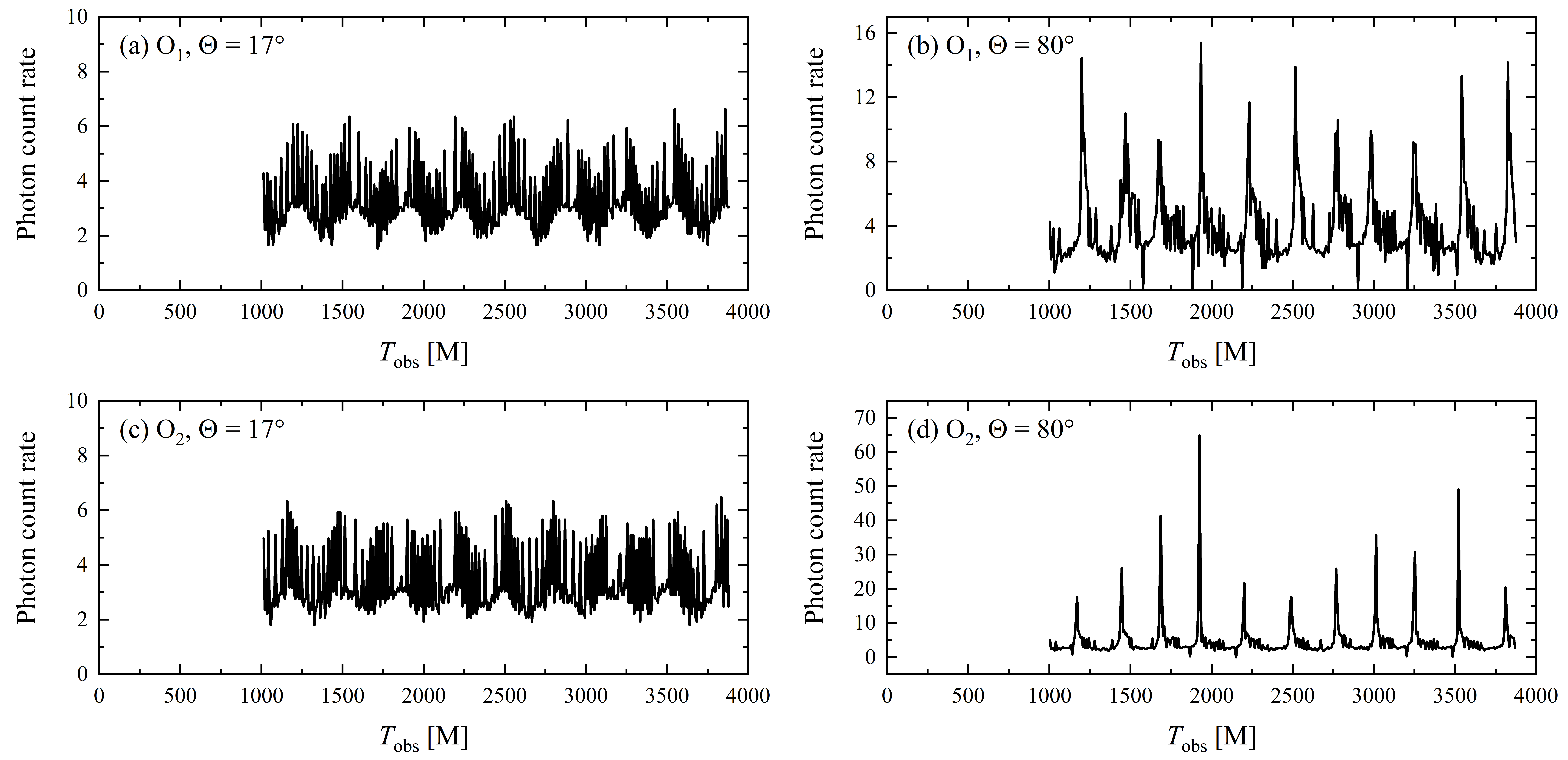}
\caption{Simulated light curves corresponding to orbits $\textrm{O}_{1}$ and $\textrm{O}_{2}$ at various observation inclination angles. Within a field of view of $x,y \in [-20,20]$, we deploy $2000^{2}$ rays to trace the discrete spheres along the orbits. The ray integration parameters are identical to those in figure \ref{tdimage}. For the construction of the light curves, we utilize the \texttt{astropy.timeseries} package, configuring the data into 400 temporal bins.}      
\label{fig14}                 
\end{figure*}

We discretize each of the two orbits into an ensemble of 500 small spheres, setting the source radius to $r_{\textrm{source}}=0.05$. Building upon this framework, the simulated light curves across various observation inclination angles are presented in figure \ref{fig14}. It is evident that at lower observation inclination angles, the light curves for both orbits exhibit similar morphological features, characterized by a distinct periodicity. As the observation inclination angle increases to $80^{\circ}$, the peaks of the light curves undergo a noticeable splitting. This peak separation indicates strong photon focusing, a direct consequence of the gravitational lensing effect. Meanwhile, the inherent periodicity of the light curves remains highly discernible. Furthermore, we find that the peak photon count rate of the light curve corresponding to orbit $\textrm{O}_{2}$ is significantly higher, reaching approximately four times the magnitude of that for $\textrm{O}_{1}$. This discrepancy arises because, from a near edge-on perspective, an orbit strictly confined to the equatorial plane possesses a smaller projected area on the observation screen; consequently, the effective cross-section for the discrete spheres to intercept the rays is substantially reduced. In contrast, an orbit exhibiting precession in the $\theta$-direction traverses a broader spatial volume, enabling it to intercept rays much more efficiently. Amplified by the gravitational lensing effect, this geometric advantage translates into an exceptionally high photon count rate during specific time intervals.
\subsection{Computational Efficiency}
To evaluate the computational efficiency of \texttt{SpinOctopus}, we benchmark its performance across two distinct hardware platforms. The first platform, denoted as $\textrm{D}_{\textrm{A}}$, is a standard desktop computer equipped with an Intel Core i7-14700 processor (utilizing 28 concurrent threads), 16 GB of RAM, and running the Windows 11 operating system. The second platform, denoted as $\textrm{D}_{\textrm{B}}$, is a high-performance workstation featuring dual AMD EPYC 7763 processors (providing 256 concurrent threads), 64 GB of RAM, and operating on a Linux system. Across both platforms, the source code is compiled using the \texttt{gfortran} compiler.
\begin{table*}
\centering
\caption{Execution times (in seconds) for simulating geometrically thick accretion disk images at various resolutions on hardware platforms $\textrm{D}_{\textrm{A}}$ and $\textrm{D}_{\textrm{B}}$. Here, a conservative adaptive step-size strategy is adopted. See the main text for details on unlocking higher computational efficiency without sacrificing precision.} 
\label{tab2}
\begin{tabular}{cccccc}
\hline
\hline
Resolution & $100^{2}$ & $500^{2}$ & $1000^{2}$ & $2000^{2}$ & $4000^{2}$  \\
\hline
$\textrm{D}_{\textrm{A}}$ & $0.8$ & $23.1$ & $101.0$ & $418.7$ & $1695.3$ \\
$\textrm{D}_{\textrm{B}}$ & $0.3$ & $3.3$ & $13.0$ & $51.3$ & $203.1$ \\
\hline
\end{tabular}
\end{table*}

Adopting the parameters of Case$_{\textrm{B}}$ from table \ref{tab1} and fixing the integration parameters to $h_{\textrm{min}}=0.001$, $N=1.8$, and $\xi_{\theta}=0.01$, with an RKF56 tolerance of $10^{-8}$, we simulate images of the geometrically thick accretion disk at an observation inclination of $\Theta=17^{\circ}$ across various resolutions. The corresponding execution times are documented in table \ref{tab2}. It is observed that at the mainstream resolution of $1000 \times 1000$ pixels, platform $\textrm{D}_{\textrm{A}}$ completes the task in a mere 101 seconds, achieving a ray-tracing efficiency of 353 rays per second per thread. Under identical conditions, platform $\textrm{D}_{\textrm{B}}$ yields a thread-level efficiency of 300 rays per second per thread. Meanwhile, it is verified that all images generated for table \ref{tab2} are entirely free of anomalous intensity pixels. 

As dictated by our adaptive step-size strategy in equation \eqref{17}, the specific combination of the parameters $h_{\textrm{min}}$, $N$, and $\xi_{\theta}$ significantly influences the trade-off between computational precision and execution efficiency. For instance, owing to the numerical singularities near the spin axis, it is theoretically necessary to employ $\xi_{\theta}$ to further reduce the integration step-size, thereby averting the emergence of anomalous pixels. However, an excessively large $\xi_{\theta}$ inevitably decelerates the ray integration process in regions away from the poles. In other words, for varying physical setups and simulation tasks, there exists an optimal combination of integration parameters that perfectly balances accuracy and performance. 

In reality, the benchmark results presented in table \ref{tab2} represent a conservative testing scenario. Depending on the target resolution, we can further decrease $\xi_{\theta}$ or increase $h_{\textrm{min}}$ to fully unlock the efficiency potential of the code. As an illustration, for the $1000 \times 1000$ resolution, if we adopt $h_{\textrm{min}}=0.0025$, $N=1.8$, and $\xi_{\theta}=0.01$ (consistent with figure \ref{fig4}), the execution time on platform $\textrm{D}_{\textrm{A}}$ drops drastically to 44.0 seconds, translating to an outstanding computational efficiency of 811 rays per second per thread. Even under these optimized conditions, the ray-tracing results remain thoroughly robust, with absolutely no anomalous pixels detected across the entire observation screen.
\section{Conclusion and Discussion}
In this paper, we introduced \texttt{SpinOctopus},\footnote{The \texttt{SpinOctopus} codebase is now publicly available at \url{https://github.com/Shiyang-Hu/Ray-Tracing-Code-SpinOctopus}.} a versatile relativistic ray-tracing algorithm and multi-functional application framework designed for arbitrary axisymmetric spacetimes, powered by a Fortran/Python hybrid architecture and OpenMP multi-threading acceleration. At its core, the framework employs a RK6 scheme for integrating timelike geodesics and an RKF56 integrator for null geodesics. Benefiting from our innovative adaptive step-size formula \eqref{17}---which is constructed based on the radial coordinate and polar angle of the propagating rays---the computational efficiency for simulating geometrically thick accretion disk images reaches an impressive $\sim 800$ rays per second per thread on a standard desktop platform ($\textrm{D}_{\textrm{A}}$).

In its current iteration, \texttt{SpinOctopus} supports the simulation of black hole critical curves, null and timelike orbits, and their corresponding light curves. Furthermore, it boasts robust imaging capabilities across a diverse array of astrophysical source models, including geometrically thick and thin accretion disks, bipolar jets, tilted thin disks, point sources, and spherical shell emitters. To execute a simulation, users are solely required to define the covariant metric tensor of the target spacetime and specify the task parameters within a dedicated configuration file, thereby completely eliminating the steep learning curve and the burden of debugging or modifying the underlying source code. Regarding extensibility, beyond the built-in models presented in this work, \texttt{SpinOctopus} provides transparent programming interfaces that allow users to seamlessly define customized source geometries and kinematic profiles. In summary, offering a rich repertoire of simulation tasks, exceptional computational efficiency, user-friendly operation, and robust extensibility, \texttt{SpinOctopus} stands as a powerful numerical toolkit for researchers in high-energy astrophysics, gravitational physics, and related disciplines.

Looking ahead, we plan to further expand the capabilities of \texttt{SpinOctopus}. Future upgrades will include modules for simulating Fe K${\alpha}$ fluorescent line profiles, geometrically thick tilted accretion disks, and the gravitational wave emission originating from timelike orbits. We also aim to establish comprehensive compatibility with discrete numerical spacetimes. Furthermore, porting the codebase to the CUDA framework for GPU acceleration and developing a Graphical User Interface (GUI) are among our primary developmental objectives.

\acknowledgments
This work was supported by the National Natural Science Foundation of China under Grant Nos. 12403081 and 12661090. For data visualization, we utilize the following Python libraries: \texttt{NumPy} \cite{2011CSE....13b..22V}, \texttt{Matplotlib} \cite{Hunter:2007}, \texttt{Astropy} \cite{2013A&A...558A..33A}, \texttt{Sympy} \cite{sympy:2017}, and \texttt{SciPy} \cite{2020SciPy-NMeth}.
\bibliographystyle{JHEP}
\bibliography{references}
\end{document}